\documentclass[reprint,
 aps,
 pre,
 floatfix,
 nofootinbib
]{revtex4-2}
\usepackage{custom-definitions}
\usepackage[all]{nowidow}
\usepackage{cleveref}

\newcommand*\widefbox[1]{\fbox{\hspace{2em}#1\hspace{2em}}}

\begin{document}

\preprint{APS/123-QED}
\renewcommand*{\thefootnote}{\fnsymbol{footnote}}

\title{Statistical Mechanics of Support Vector Regression}

\author{Abdulkadir Canatar}
\author{SueYeon Chung}
\affiliation{
 Center for Computational Neuroscience, Flatiron Institute, New York, NY, 10010\\
 Center for Neural Science New York University, New York, NY, 10003, USA
}

\date{\today}

\begin{abstract}
A key problem in deep learning and computational neuroscience is relating the geometrical properties of neural representations to task performance. Here, we consider this problem for continuous decoding tasks where neural variability may affect task precision. Using methods from statistical mechanics,  we study the average-case learning curves for $\varepsilon$-insensitive Support Vector Regression ($\varepsilon$-SVR) and discuss its capacity as a measure of linear decodability. Our analysis reveals a phase transition in training error at a critical load, capturing the interplay between the tolerance parameter $\varepsilon$ and neural variability. We uncover a double-descent phenomenon in the generalization error, showing that $\varepsilon$ acts as a regularizer, both suppressing and shifting these peaks. Theoretical predictions are validated both with toy models and deep neural networks, extending the theory of Support Vector Machines to continuous tasks with inherent neural variability.
\end{abstract}

\maketitle

\section{Introduction}

Understanding how neural systems and machine learning models perform robust decoding in the presence of task-irrelevant variability is a fundamental challenge ~\cite{poggio2016visual, lecun2015deep, goodfellow2009measuring}. In real-world tasks, input representations often contain fluctuations that do not affect the target variable of interest. For example, estimating the position of a specific animal from video frames requires ignoring variations due to their sizes, lighting changes, and backgrounds~\cite{hong2016explicit}. Such nuisance variability complicates learning, especially when data are limited.

Robust decoding thus demands representations that preserve task-relevant features while remaining tolerant to task-irrelevant variability. In both biological and artificial systems, this tolerance can be viewed as a form of perceptual invariance—the ability to treat small deviations as equivalent for the purposes of a task. Quantifying how much variability a representation can accommodate without loss of decoding precision remains an open problem in both neuroscience and machine learning.

In this work, we introduce a theoretical framework to quantitatively characterize $\varepsilon$-insensitive Support Vector Regression ($\varepsilon$-SVR)~\cite{vapnik1996svr, scholkopf2000new} to analyze the robustness of continuous decoding. In $\varepsilon$-SVR, deviations smaller than a tolerance $\varepsilon$ are not penalized during learning, allowing us to directly model how representational variability affects decoding accuracy. The $\varepsilon$-SVR task is similar to ridgeless regression, except that the objective is satisfied when the prediction error lies within a tube of size $\varepsilon$ around the target (\figref{fig:figure1}). Here, the \textit{tube size} $\varepsilon$ acts as a hyperparameter that defines the margin of tolerance for input deviations and must be fine-tuned depending on the structure of the variability.

Let $P$ denote the number of training samples, $N$ the dimensionality of the representation, and $\alpha:=P/N$ the \textit{load}. In ridgeless regression, when input features and/or targets are noisy, the training error $E_{tr}$ goes through a phase transition, from being exactly zero to being strictly non-zero, at the critical load $\alpha_c = 1$, also known as the interpolation threshold. At this point, double-descent occurs \cite{belkin2019reconciling}; the generalization error $E_g$ diverges due to overfitting and starts improving again for $\alpha > \alpha_c$. As $\alpha\to\infty$, training and generalization errors approach each other and asymptotically become the same.

By tuning an appropriate tube size, SVR allows fitting more samples ($\alpha_c>1$) with zero training error by tolerating errors up to $\varepsilon$. Alternatively, one can think of shifting the interpolation threshold beyond $\alpha=1$ in ridgeless regression\footnote{Qualitatively, SVR still differs from ridgeless regression with a shifted interpolation threshold.}. This is always possible by picking an arbitrarily large $\varepsilon$, in which case no learning occurs. Hence, we consider the minimum $\varepsilon^*$ for which the training error is zero. This rate is interpreted as the decoding precision of the representation. In other words, the amount of tolerance $\varepsilon^*$ required to learn all training examples successfully quantifies how precisely a representation can decode the task. Furthermore, this also defines a form of invariant representation, namely $\varepsilon$-invariance, since the objective function only requires the predictions to be in the $\varepsilon$-vicinity of the exact target. 

Using the replica method \cite{mezard1987spin,engel2001statistical,mezard2009information}, we develop an analytical theory of generalization for $\varepsilon$-SVR that relates the precise geometric properties of representations to their decoding precision of continuous variables. Our theory recovers the previous results on ridge regression in the limit $\varepsilon \to 0$ \cite{bordelon2020spectrum,jacot2020implicit,canatar2021spectral,simon2023eigenlearning,atanasov2022onset,atanasov2024scaling}. When applied to our setting, our results also agree with the predictions of the theory by \citet{loureiro2021learning} developed for generic loss functions (see SI.\ref{sec:SI_loureiro}).

The rest of the paper is organized as follows:

In \secref{sec:background}, we review the relevant literature and discuss the similarities with our work. 

In \secref{sec:theory}, we assume Gaussian representations with arbitrary covariance structure and derive analytical expressions for the training and generalization errors of $\varepsilon$-SVR when presented with $P$ training samples. We show that, for fixed $P$, the training error undergoes a phase transition at a critical value $\varepsilon^*$ beyond which it becomes exactly zero. We introduce and numerically test a toy data model and discuss the implications of the theory.

In \secref{sec:optimal_rates}, we derive optimal learning curves for both ridge regression and support vector regression. We show that optimal ridge regression always outperforms optimal $\varepsilon$-SVR.

In \secref{sec:real_data}, we apply our theory to deep convolutional neural network representations on a continuous decoding task of grating images. We find that training neural networks on natural images allows better discriminability of angles.

\section{Background}\label{sec:background}

Understanding geometric properties of neural representations of data with invariances is a longstanding problem in both machine learning \cite{bengio2013representation,achille2018emergence} and neuroscience \cite{chung2021neural}. Here, we review two main lines of prior work related to invariant decoding.

One line of work studies how known symmetries of a particular task can be exploited in designing tailored machine learning models \cite{mei2021learning, favero2021locality, elesedy2021provably}. These works demonstrate that regression with kernels that are explicitly invariant under the group actions of the symmetry group of some task significantly improves the sample efficiency. These representations express only task-relevant input variances and hence effectively overcome the curse of dimensionality. Instead of group invariances, here we consider $\varepsilon$-invariance in generic representations, quantifying the amount of decoding precision that must be sacrificed to achieve zero training error. One can also imagine simplifying the decoding problem and discretizing the output variable in a way that input variances do not affect the performance anymore.

A second line of research examines how geometric properties of neural representations relate to invariant prediction, with a particular emphasis on linear decodability under neural variability~\cite{chung2018classification, chung2021neural, wakhloo2023linear}. These studies analyzed classification performance using Support Vector Classification (SVC), linking the number of separable object manifolds to the geometry of the representation. In related work,~\citet{farrell2021capacity} explored how continuous group invariances influence classification capacity. In this paper, we extend this line of inquiry to continuous regression tasks.

\section{Theoretical Framework}\label{sec:theory}
We consider a supervised learning setting in which inputs are represented by $N$-dimensional \textit{center} representations $\bar\bpsi$, and labels $\bar y$ are linearly generated via a coding direction $\bar\w$, i.e. $\bar y = \bar\w\cdot\bar\bpsi$. 
Here, the center representations $\bar\bpsi$ are treated as random variables and capture the task-relevant information from the input data.
\begin{figure}[ht]
    \centering    
    \includegraphics[width=.8\linewidth]{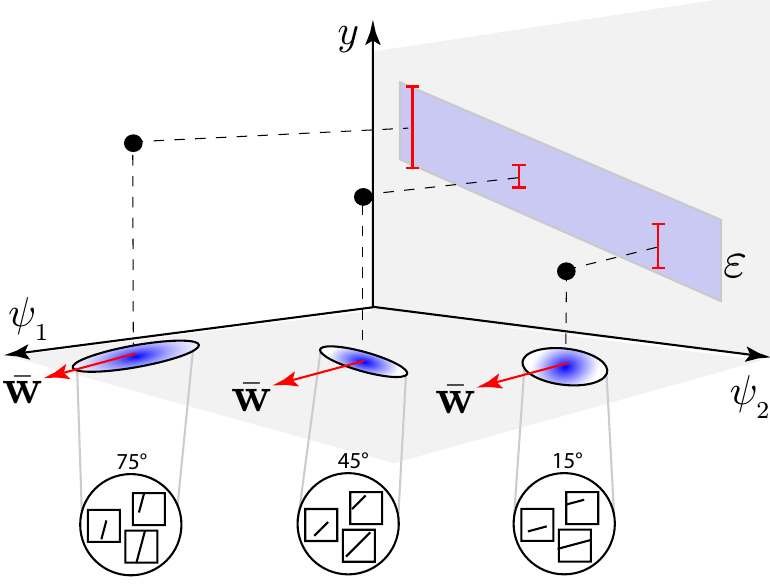}
    \caption{Decoding the orientation of sticks (degrees) from their pixel representations as an example of invariant regression. For each orientation, labelled circles depict stick images with different variations, such as location and size.  Blue regions depict the corresponding neural representations in $(\psi_1, \psi_2)$-space induced by task-irrelevant variances and depict possible noise geometries with $\bdelta$ parallel (left), orthogonal (middle), and partially aligned (right) with the coding direction $\bar\w$.}
    \label{fig:figure1}
\end{figure}

In contrast, the learner observes \textit{encoding} representations $\bpsi$ which include both the task-relevant centers $\bar\bpsi$ and task-irrelevant variations modeled by random variables $\bdelta$, and assumed to be of the form $\bpsi \equiv \bar\bpsi + \bdelta$. The \textit{noise} representations $\bdelta$ represent nuisance variations around the centers. The goal is to learn a linear predictor $y = \w \cdot \bpsi$. Similar data models have also been analyzed in prior studies of ridge regression \cite{atanasov2022onset,atanasov2024scaling,loureiro2021learning,ruben2023learning,adlam2020neural}.

For each training set instance, $P$ samples $\{\bar\bpsi^\mu, \bpsi^\mu\}_{\mu=1}^P$ are drawn i.i.d. from their respective distributions, and used to generate labels $\bar y^\mu$ and predictions $y^\mu$. Due to the additional noise, the predictions always mismatch with the labels when $P>N$ and overfit to noise when $P<N$. To account for this variance, we consider the $\varepsilon$-SVR algorithm
\begin{align}\label{eq:svr_problem}
    \min_{\w \in \bR^N} \frac{1}{2} \norm{\w}^2 + \frac{1}{2\lambda}\sum_{\mu=1}^P \lrsqpar{\abs{y^\mu-\bar y^\mu} - \varepsilon}_+^2
\end{align}
where $\varepsilon$ is the tube size, $\lambda$ is the ridge parameter and $\lrsqpar{x}_+ \equiv \max(0, x)$. We note that both $\lambda$ and $\varepsilon$ serve as regularizers for regression since both control the effect of training error on the solution. However, we later show that they have qualitatively different effects on generalization (see SI.\ref{sec:SI_replica}). Although our analysis holds for general $\lambda$ and $\varepsilon$, here we focus on the ridgeless limit ($\lambda = 0$), where the objective is fitting all training samples exactly within the $\varepsilon$-tube.

Our goal is to compute the average-case training and generalization errors for $\varepsilon$-SVR. By treating the training set as a quenched disorder, we compute these average quantities using the replica method from statistical physics and consider the following partition function:
\begin{align}\label{eq:partition_func}
    &\overline{\log Z} = \overline{\log \int d \w \, e^{-N S(\w, \bpsi^\mu, \bar\bpsi^\mu) + J\cO(\w)}}\\
     &S = \frac{1}{2N}\norm{\w}^2 + \frac{1}{2\lambda N} \sum_{\mu=1}^P \lrsqpar{\abs{\w\cdot \bpsi^\mu - \bar\w\cdot \bar\bpsi^{\mu}}-\varepsilon}_+^2,\nonumber
\end{align}
where the overline denotes the quenched average over the training set $\{\bpsi^\mu, \bar\bpsi^\mu\}$ of size $P$. In the thermodynamic limit $P,N \to \infty$ while keeping $\alpha \equiv P/N \sim \cO(1)$, the free energy $S$ in Eq.\eqref{eq:partition_func} becomes self-averaging, concentrating around the solution to the problem in Eq.\eqref{eq:svr_problem}. Then, the average of any observable $\braket{\cO(\w)}$ can be computed by taking derivatives of $\overline{\log Z}$ with respect to source $J$. 

The training error $E_{tr}$ and generalization error $E_g$, the two observables we are interested in, are given by
\begin{align}\label{eq:tr_gen_err}
    E_{tr} &= \overline{\frac{1}{P}\sum_{\mu=1}^P \lrsqpar{\abs{\w\cdot \bpsi^\mu - \bar\w\cdot \bar\bpsi^{\mu}}-\varepsilon}_+^2}\nonumber\\
    E_g &= \overline{\Braket{\lrpar{\w\cdot \bpsi - \bar\w\cdot \bar\bpsi}^2}_{\bpsi,\bar\bpsi}}.
\end{align}

\subsection{Replica Calculation}

For simplicity, we assume a joint Gaussian distribution over center and encoding representations:
\begin{align}
    \begin{bmatrix}\bpsi \\ \bar\bpsi\end{bmatrix} \sim \cN\lrpar{0, \begin{bmatrix} \bSigma_\bpsi & \bSigma_{\bpsi\bar\bpsi} \\ \bSigma_{\bpsi\bar\bpsi}^\top & \bSigma_{\bar\bpsi}\end{bmatrix}}.
\end{align}
Note that the solution to the objective in Eq.\eqref{eq:partition_func} under this data distribution depends critically on the correlation between centers $\bar\bpsi$ and encoding representations $\bpsi$. Specifically, in the large-sample limit $P\to\infty$, the optimal weights and generalization error converge to:
\begin{align}
    E_{\infty} = \tr \bar\w\bar\w^\top\bSigma_{\bar\bpsi} -  \tr \w_*\w_*^\top\bSigma_{\bpsi},\; \w_* = \bSigma_\bpsi^{-1}\bSigma_{\bpsi\bar\bpsi}\bar\w,\nonumber
\end{align}
where $\tr$ denotes the normalized trace operator. The vector $\w_*$ represents the projection of the target weights $\bar\w$ into the subspace accessible to the learner, and $E_\infty$ denotes the irreducible error corresponding to the unexplainable variance in the target \cite{atanasov2022onset}.

By performing the quenched average using this distribution—under a replica symmetric ansatz and saddle-point approximation (see SI.\ref{sec:SI_replica})—we obtain a mean-field theory characterized by three effective parameters: the \textit{effective regularization} $\tilde\lambda$, the \textit{effective tube size} $\tilde\varepsilon$, and the \textit{effective load} $\tilde\alpha$:
\begin{align}\label{eq:self_consistent}
    \tilde \lambda &=  \lambda + \cF(\tilde \lambda, \tilde\varepsilon), \quad \tilde\varepsilon = \varepsilon \big /\sqrt{\cG(\tilde \lambda, \tilde\varepsilon)}, \quad \tilde \alpha = \alpha f\lrpar{\tilde\varepsilon}
\end{align}
where $\tilde \lambda$ and $\tilde \varepsilon$ satisfy the following coupled self-consistent equations:
\begin{align}
    \cF(\tilde \lambda, \tilde\varepsilon) &= \tr \M, \;\; \M = \bSigma_\bpsi \G^{-1}, \;\; \G = \I + \frac{\tilde \alpha}{\tilde\lambda} \bSigma_\bpsi \nonumber\\
    \cG(\tilde \lambda, \tilde\varepsilon) &= \frac{E_\infty + \tr \M \G^{-1}\w_*\w_*^\top}{1- g(\tilde\varepsilon) \gamma}, \;\; \gamma = \partial_{\tilde\lambda}\cF(\tilde \lambda, \tilde\varepsilon).
\end{align}
Here, the scalar functions $f(\tilde\varepsilon)$ and $g(\tilde\varepsilon)$ are given by:
\begin{align}
    f(\tilde\varepsilon) &= \Braket{\theta\lrpar{\abs{z} - \tilde\varepsilon}}_z, \;\; g(\tilde\varepsilon) = \frac{\Braket{\lrpar{\abs{z} - \tilde\varepsilon}^2\theta\lrpar{\abs{z} - \tilde\varepsilon}}_z}{\Braket{\theta\lrpar{\abs{z} - \tilde\varepsilon}}_{z}}\nonumber
\end{align}
where $\braket{f(z)}_z$ denotes expectation with respect to a standard normal variable $z$. Furthermore, the function $\cG(\tilde \lambda, \tilde\varepsilon)$ coincides with the generalization error $E_g$ from Eq.~\eqref{eq:tr_gen_err}, and the training error is given by: 
\begin{align}
    E_{tr} &= \lrpar{1 - \frac{\cF(\tilde \lambda, \tilde\varepsilon)}{\tilde\lambda}}^2 f(\tilde\varepsilon) g(\tilde\varepsilon) E_{g}, \quad E_{g} = \cG(\tilde \lambda, \tilde\varepsilon).
\end{align}
The effective regularization $\tilde\lambda$ is a renormalized ridge parameter \cite{cheng2022dimension, atanasov2024scaling} that governs the smoothness of the solution and quantifies the implicit regularization of the model \cite{jacot2020implicit,canatar2021spectral}.

The effective load $\tilde\alpha$ gives a measure of the actual number of samples that contribute to learning, and decreases monotonically as a function of the effective tube size $\tilde\varepsilon$. Intuitively, larger tube sizes require sampling more points since it becomes less likely to hit the tube boundary. A similar notion was also introduced in \cite{bos1993generalization} when the learning is restricted to a subset of training examples.

The effective tube size $\tilde\varepsilon$, on the other hand, is a renormalized measure of tolerance and depends inversely on generalization error, implying that the number of samples required to improve generalization increases as the model's predictivity improves (Eq.\eqref{eq:self_consistent}).

Finally, $\tilde\varepsilon = {\varepsilon}/{\sqrt{E_g}}$ can also be interpreted as a form of effective discriminability varying as a function of load $\alpha$. In neuroscience, \textit{discriminability} $d' \equiv {\varepsilon}/{\sqrt{E_\infty}}$ is a common metric used to quantify the perceptible contrast between two stimuli $\varepsilon$-apart from each other \cite{seung1993simple, brunel1998mutual}.

\subsection{Analysis of a Toy Data Model}

To analyze our results and build intuition, we apply our theory to investigate the decoding precision in the presence of structured noise, which correlates with the coding direction. We consider a toy data model where the centers are drawn from $\bar\bpsi \sim \cN(0,\I)$ and the labels are generated by $\bar y = \bar\w\cdot\bar\bpsi$ with $\tr\bar\w\bar\w^\top = 1$. The representations seen by the linear model are of the form $\bpsi = \bar\bpsi + \sigma\bdelta$ where the noise $\bdelta$ is drawn from a structured Gaussian distribution with zero mean and covariance
\begin{align}
    \bSigma_\bdelta &= (1-\beta)\P + \beta (\I - \P),\quad \P \equiv \I - \frac{1}{N}\bar\w\bar\w^\top,
\end{align}
where $\sigma$ controls the noise strength. Through the parameter $\beta$, this noise model interpolates between the cases where the noise is completely orthogonal ($\beta=0$) or parallel ($\beta=1$) to the coding direction $\bar\w$. A schematic of this data model is illustrated in \figref{fig:figure1}.

\begin{figure}[ht]
    \centering
    \includegraphics[width=.95\linewidth]{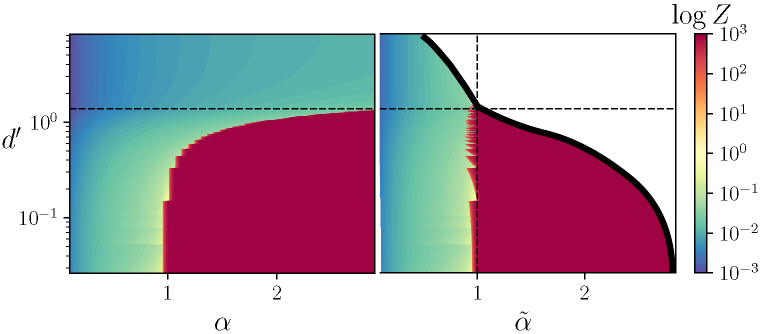}
    \caption{Phase plots of the theoretical values of $\overline{\log Z}$ against $d' = \varepsilon/ \sqrt{E_\infty}$. (Left) For fixed $E_\infty$, increasing $\varepsilon$ delays phase transition, and for fixed $\varepsilon$, increasing noise variability ($\sigma$) causes a phase transition for smaller sample sizes. (Right) Same plot but as a function of $\tilde\alpha$. The vertical dashed line indicates the phase boundary at $\tilde\alpha = 1$, the thick solid curve is the boundary $\alpha = \alpha_{max}=3$, and the horizontal line indicates their intersection corresponding to the minimum $d'$ to achieve zero training error.}
    \label{fig:figure2}
\end{figure}

For this model, the training and generalization errors from Eq.\eqref{eq:tr_gen_err} simplify to the following forms (see SI.\ref{sec:SI_toy_data_model}):
\begin{align}\label{eq:train_gen_error_toy}
    E_g &= \frac{1}{1-\frac{g(\tilde\varepsilon)\tilde\alpha}{(\tilde\alpha+\tilde\lambda)^2}}\lrpar{E_\infty + \frac{1-E_\infty}{\lrpar{1 + \frac{\tilde\alpha}{\tilde\lambda}\frac{1 + \beta\sigma^2}{1 + (1-\beta)\sigma^2}}^2}}\nonumber\\
    E_{tr} &= \lrpar{1-\frac{1}{\tilde\alpha+\tilde\lambda}}^2f(\tilde\varepsilon)g(\tilde\varepsilon)E_g, \quad E_\infty = \frac{\beta\sigma^2}{1+\beta\sigma^2},
\end{align}
where the self-consistent equations for $\tilde\lambda$ and $\tilde\varepsilon$ in Eq.\eqref{eq:self_consistent} can be solved numerically. Furthermore, in the ridgeless limit $\lambda\to 0$, the expression for $\tilde\lambda$ simplifies to:
\begin{align}\label{eq:effective_alpha_solution}
    \tilde\lambda = 
    \begin{cases}
        1 - \tilde\alpha, & \tilde\alpha < 1\\
        0, & \tilde\alpha \geq 1.
    \end{cases}
\end{align}
Eq.\eqref{eq:train_gen_error_toy} and Eq.\eqref{eq:effective_alpha_solution}  imply that in the over-parameterized regime, when the effective sample size is smaller than the number of parameters ($\tilde\alpha < 1$), the training error vanishes identically and goes through a first-order phase transition at $\tilde\alpha = 1$ corresponding to the interpolation threshold. Beyond this threshold ($\tilde\alpha > 1$), $E_{tr}$ is non-zero and approaches the generalization error for large $\tilde\alpha$. Furthermore, when $\tilde\alpha>1$, the generalization error $E_g$ becomes proportional to $E_\infty$, and thus identically vanishes when either $\beta$ or $\sigma$ is zero. In \figref{fig:figure2}, we visualize this phase transition by plotting the free energy across model parameters.

In the limit $\varepsilon \to 0$, our theory recovers the equations from previous work on ridge regression \cite{jacot2020implicit, canatar2021spectral, adlam2020neural}. However, we discover additional phenomena in SVR that are lacking in ridge regression and are reported below.

\subsection{Model Capacity and Effective Sample Size}

\begin{figure}[ht]
    \centering
    \includegraphics[width=.99\linewidth]{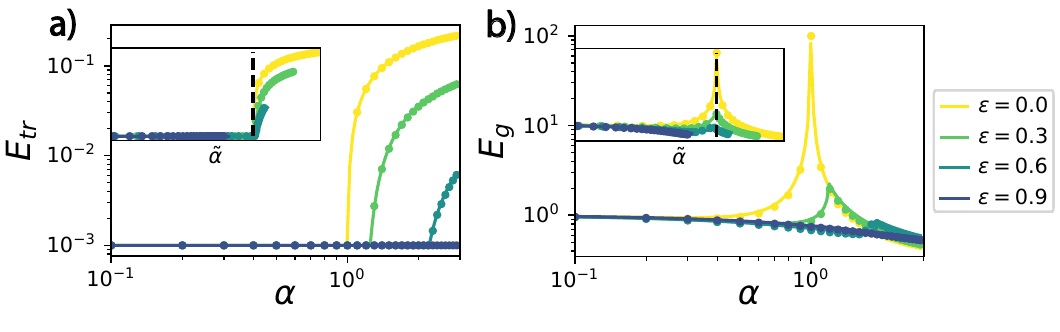}
    \caption{Theoretical (lines) and experimental (dots) learning curves for the toy model with $\beta=0.5$ and $\sigma=1$. Insets show the same curves but as a function of $\tilde\alpha$. \textbf{a)} With increasing $\varepsilon$-tube size, training error decreases due to increasingly relaxed definition of being correct, therefore allowing more samples to be fit. \textbf{b)} On the other hand, generalization error displays a non-monotonicity called the double-descent. In this case, the effect of increasing $\varepsilon$ amounts to keeping the problem in the \textit{overparameterized} regime.}
    \label{fig:figure3}
\end{figure}

Our analysis reveals that the phase boundary is governed by the effective load $\tilde\alpha = 1$ (see \figref{fig:figure2}). This critical point is clearly observed for both training error $E_{tr}$ and generalization error $E_g$ in \figref{fig:figure3}, where the theoretical predictions closely match empirical results.

In ridge regression ($\varepsilon = 0$), the training error undergoes a phase transition at $\alpha = 1$. For SVR with $\varepsilon > 0$, the transition shifts to a higher value $\alpha > 1$ (Fig.~\ref{fig:figure3}a). This is expected since larger tube sizes allow for more \textit{capacity} in the sense that more samples can be fit satisfying $E_{tr} = 0$. Hence, the transition for non-zero $\varepsilon$ occurs at some critical load $\alpha_c > 1$
\begin{align}\label{eq:critical_epsilon}
    \alpha_c^{-1} = f(\tilde\varepsilon)
\end{align}
so that $\tilde\alpha = 1$ (see \figref{fig:figure3}a, inset). This critical value $\alpha_c$ represents the maximum number of training examples that can be fit within an $\varepsilon$-tube and, in analogy to classification capacity \cite{chung2018classification}, can be used as a measure that ties representational geometry to the precision of continuous decoding tasks.

Computing the capacity $\alpha_c$ requires solving the coupled equations in Eq.\eqref{eq:self_consistent} which are, in general, analytically intractable. However, it can be exactly solved for our toy model Eq.\eqref{eq:train_gen_error_toy} at the critical point $\tilde\alpha = 1$ (see SI.\ref{sec:SI_toy_data_model}). In this case, the solution asymptotically has the form:
\begin{align}
    \tilde\varepsilon \approx \begin{cases}
        d' - 1/d', & d' \gg 1 \\
        d'^{2} \sqrt{\frac{2}{\pi}}, & d' \ll  1
    \end{cases},
\end{align}
Through Eq.\eqref{eq:critical_epsilon}, this result explicitly links the capacity $\alpha_c$ to discriminability $d'$, and shows that lower discriminability (larger tube sizes) permits fitting more training samples without error.

Furthermore, it allows us to connect the model's capacity to its representational properties since $d'$ explicitly depends on model parameters $\sigma$ (noise) and $\beta$ (task correlation). For fixed tube size $\varepsilon$, discriminability $d'$ increases with more noise and task correlation. On the other hand, when the noise geometry is completely orthogonal to the task ($\beta=0$), the noise magnitude does not affect $d'$ and hence the capacity.

\subsection{Generalization and Double-Descent}\label{sec:generalization}

The analysis of $E_g$ reveals a well-known phenomenon known as \textit{double-descent} \cite{belkin2018understand,belkin2019reconciling} where the generalization degrades as the number of samples approaches the critical point $\tilde\alpha = 1$ (\figref{fig:figure3}b).  Essentially, double-descent implies overfitting to noisy samples at the model capacity and signals a sharp transition between the over- and under-parameterized regimes.

In regression, this overfitting can be avoided by tuning the ridge parameter $\lambda$, which penalizes solutions with large norms \cite{scholkopf2002learning}. In our case, although we set the ridge parameter to zero, turning on $\varepsilon$ also acts like a regularization. However, in contrast to the ridge penalty, which only suppresses the double-descent peak, $\varepsilon$ additionally shifts its location, implying that the tube size effectively controls whether the model operates in an over- or under-parameterized regime (\figref{fig:figure3}b).

Similar behavior has been previously observed in many ridge-regression settings where increasing the number of effective parameters of a model helps shift the double-descent peak \cite{maloney2022solvable,atanasov2022onset,ruben2023learning}. Based on these observations, we may also consider $\varepsilon$ as a parameter that tunes the effective model size.

\section{Optimal Ridge Regression vs. Optimal Support Vector Regression}\label{sec:optimal_rates}

Determined by the critical point in training error, the capacity condition in Eq.\eqref{eq:critical_epsilon} determines either the maximum allowable load $\alpha$ for a fixed tolerance $\varepsilon$, or the minimum possible tolerance $\varepsilon$ required to fit a dataset of size $\alpha$. However, at the critical point ($\tilde\alpha = 1$), the generalization error typically diverges (\figref{fig:figure2}b), implying that the model completely overfits at capacity with poor generalization. For this reason, we define the \textit{optimal} load $\alpha_{\text{opt}} < \alpha_c$ as the maximum number of samples that can be fit with zero training error while maintaining acceptable generalization.

Here, we analyze this quantity by minimizing the generalization error in Eq.\eqref{eq:tr_gen_err} with respect to model hyperparameters. Treating $\varepsilon$ as a free parameter, we compute the optimal tolerance $\varepsilon_{opt}$ as a function of sample size and obtain analytical results for optimal $\varepsilon$-SVR. For comparison, we also derive learning curves for optimal ridge regression by minimizing $E_g$ with respect to $\lambda$ \cite{wu2020optimal} and compare their generalization performances. Since we get analytical expressions for both $\varepsilon_{opt}$ and $\lambda_{opt}$, our results furthermore help bypass computationally intensive hyperparameter sweeps for ridge regression and $\varepsilon$-SVR \cite{hastie2009elements}.

For optimal SVR, we set $\lambda = 0$ and solve for $\partial_\varepsilon E_g = 0$. While the analytical calculation for optimal $\varepsilon$ is tedious (see SI.\ref{sec:SI_optimal_learning_rates}), we can obtain a new set of self-consistent equations replacing the ones in Eq.\eqref{eq:self_consistent}. For optimal SVR, we get the following self-consistent equations:
\begin{align}
\tilde \lambda &= \tr\M, \quad h({\tilde \varepsilon}_{opt}) =  \frac{\cA({\tilde \varepsilon}_{opt}, {\tilde \lambda})}{1-\gamma},
\end{align}
where 
\begin{align}
    \cA({\tilde \varepsilon}, {\tilde \lambda})&= \frac{\tr\M^2\G^{-1}}{\tr\M^2}\lrpar{g(\tilde\varepsilon) - 
 \frac{{\tilde \lambda}}{C}\frac{\tr\M^2\G^{-1} {\w_*}{\w_*}^\top}{\tr\M^2\G^{-1}}},\nonumber\\
     h(\tilde\varepsilon) &\equiv \tilde\varepsilon\lrpar{\tilde\varepsilon + \frac{f'}{f} - \frac{f}{f'}}.
\end{align}
While being complicated, these self-consistent equations can be easily solved numerically.

For optimal ridge regression, we set $\varepsilon = 0$ and solve for $\partial_\lambda E_g = 0$. In this case, the self-consistent equation simply reduces to $\cA({\tilde \varepsilon}, {\tilde \lambda}) = 0$ (see SI.\ref{sec:SI_optimal_learning_rates}) and yields:
\begin{align}
    {\tilde \lambda} =  C\frac{\tr\M^2\G^{-1}}{\tr\M^2\G^{-1} {\w_*}{\w_*}^\top}
\end{align}
replacing the original equation $\tilde\lambda = \lambda + \tr \M$ in Eq.\eqref{eq:self_consistent}.

\begin{figure}[ht]
    \centering
    \includegraphics[width=.8\linewidth]{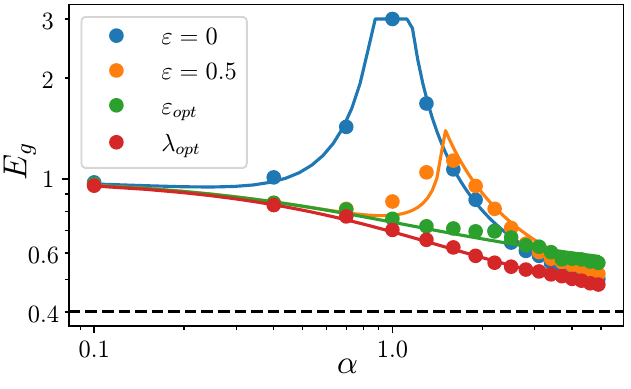}
    \caption{Optimal learning curves for the toy model. Choosing optimal $\lambda$ for ridge regression always outperforms optimal $\varepsilon$-SVR. Lines indicate theory, and dots indicate experiments. The dashed line indicates $E_\infty$. Optimal $\lambda$ and $\varepsilon$ change as a function of $\alpha$, and their values are reported in \figref{fig:SI_optimal_lambda_epsilon}.}
    \label{fig:figure4}
\end{figure}

In \figref{fig:figure4}, we test our theory and observe perfect agreement with experiments. Interestingly, optimal ridge regression consistently outperforms optimal $\varepsilon$-SVR in terms of generalization. Note that the optimal ridge regression always maintains a non-zero training error, while optimal SVR ensures zero training error for all $\alpha$. Here, optimal SVR yields worse generalization performance because it stops learning as soon as the constraints are satisfied, while in optimal ridge regression, the model is forced to learn finer details since the training error never vanishes.

\section{Real Data Applications}\label{sec:real_data}

Although our theory is derived in the self-averaging limit $N\to\infty$ and $\alpha \sim \cO(1)$, we find that it agrees well with experiments when it is applied to real data with finite $P$ and $N$.

To test the applicability of our theory, we designed an experiment where the task is to decode the orientation of grating images from their deep neural network representations. We generated grating images with orientations $\theta$ uniformly sampled from $[0, \pi)$ and augmented them with task-irrelevant attributes such as spatial frequency and phase (see SI.\ref{sec:SI_experimental_details}).
We extracted their neural network representations from the hidden layers of various ResNet architectures and evaluated their decoding performance using optimal $\varepsilon$-SVR.

\begin{figure}[ht]
    \centering
    \includegraphics[width=.99\linewidth]{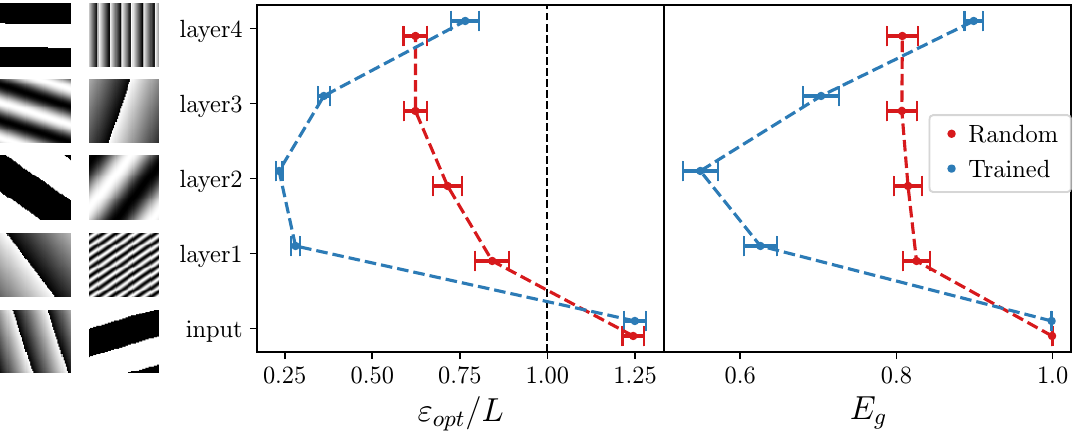}
    \caption{SVR analysis of layers extracted from trained and random ResNet models on the task of decoding grating orientations.}
    \label{fig:figure5}
\end{figure}

In \figref{fig:figure5}, we show the optimal $\varepsilon_{opt}$ and generalization error for both randomly initialized models and trained models on natural scenes. Here, $\varepsilon_{opt}$ is normalized by the range of the decoding interval $L$ such that the case $\varepsilon_{opt}/L > 1$ implies that decoding any orientation is impossible while fitting all training examples. For random networks, discriminability improves with depth. In contrast, for trained networks, intermediate layers yield better discriminability, which then degrades in deeper layers. This trend—where early and middle layers better support decoding of simple stimuli—has also been observed in prior work~\cite{hong2016explicit}.

Our theoretical framework allows the decomposition of generalization error into interpretable spectral components, offering insights into how and why decoding performance evolves across network layers. In SI.\ref{sec:SI_spectral_decomposition}, we demonstrate that the error mode geometry framework from~\cite{canatar2024spectral} can be directly applied to our setting (see SI.\ref{sec:SI_additional_experiments} for error mode geometry analysis).

\section{Discussion}

Invariant decoding remains a fundamental challenge in both machine learning and neuroscience, as real-world inputs frequently vary in ways that do not alter the target attribute. This study bridges the gap between representational geometry and its influence on continuous decoding tasks. By utilizing $\varepsilon$-SVR as a geometric probe in high-dimensional feature spaces, we employed average-case analysis to explore how the geometry of noise representations impacts decoding performance.

In particular, our analysis revealed that the phase transition in training error can be controlled by the tube size parameter $\varepsilon$, which also determines the desired precision of the task. Larger $\varepsilon$ values increase the model's capacity by allowing more samples to be fit without error, but this comes at the cost of reduced precision in continuous variable decoding, as indicated by increased generalization error. This trade-off between training and generalization errors underscores the importance of tuning $\varepsilon$ to optimize both model fit and predictive accuracy.

We also derived analytical expressions for optimal learning curves for both $\varepsilon$-SVR and ridge regression. 
While prior work also derived analytical formulas for the optimal ridge parameter~\cite{thrampoulidis2015regularized,dobriban2018high,wu2020optimal,kobak2020optimal,hastie2022surprises,zhang2024neural}, these results were limited to simplified data designs. Here, our framework allowed us to extend these results to generic Gaussian data with arbitrary covariance and compute optimal hyperparameters analytically for both ridge strength $\lambda$ and tube width $\varepsilon$ without a hyperparameter search.

While our analysis currently relies on Gaussian assumptions and linear decoding strategies, future investigations should extend beyond these assumptions and address more realistic coding scenarios in distributed neural representations.

Our theoretical framework reveals fundamental tradeoffs between precision and robustness in neural computations. The relationship we uncovered between representational geometry and decoding performance not only enhances our understanding of SVR algorithms but could also inform the design of more robust regression models and understand the nature of neural decoding algorithms where precise representations must be maintained despite inherent variability.

\begin{acknowledgments}
The authors thank Chi-Ning Chou for helpful comments on the manuscript. This work was supported by the Center for Computational Neuroscience at the Flatiron Institute of the Simons Foundation, as well as by a Sloan Research Fellowship and a Klingenstein-Simons Award (to S.C.). All experiments were performed on the high-performance computing cluster at the Flatiron Institute. 
\end{acknowledgments}

\bibliography{bibliography}

\begin{thebibliography}{47}%
\makeatletter
\providecommand \@ifxundefined [1]{%
 \@ifx{#1\undefined}
}%
\providecommand \@ifnum [1]{%
 \ifnum #1\expandafter \@firstoftwo
 \else \expandafter \@secondoftwo
 \fi
}%
\providecommand \@ifx [1]{%
 \ifx #1\expandafter \@firstoftwo
 \else \expandafter \@secondoftwo
 \fi
}%
\providecommand \natexlab [1]{#1}%
\providecommand \enquote  [1]{``#1''}%
\providecommand \bibnamefont  [1]{#1}%
\providecommand \bibfnamefont [1]{#1}%
\providecommand \citenamefont [1]{#1}%
\providecommand \href@noop [0]{\@secondoftwo}%
\providecommand \href [0]{\begingroup \@sanitize@url \@href}%
\providecommand \@href[1]{\@@startlink{#1}\@@href}%
\providecommand \@@href[1]{\endgroup#1\@@endlink}%
\providecommand \@sanitize@url [0]{\catcode `\\12\catcode `\$12\catcode
  `\&12\catcode `\#12\catcode `\^12\catcode `\_12\catcode `\%12\relax}%
\providecommand \@@startlink[1]{}%
\providecommand \@@endlink[0]{}%
\providecommand \url  [0]{\begingroup\@sanitize@url \@url }%
\providecommand \@url [1]{\endgroup\@href {#1}{\urlprefix }}%
\providecommand \urlprefix  [0]{URL }%
\providecommand \Eprint [0]{\href }%
\providecommand \doibase [0]{https://doi.org/}%
\providecommand \selectlanguage [0]{\@gobble}%
\providecommand \bibinfo  [0]{\@secondoftwo}%
\providecommand \bibfield  [0]{\@secondoftwo}%
\providecommand \translation [1]{[#1]}%
\providecommand \BibitemOpen [0]{}%
\providecommand \bibitemStop [0]{}%
\providecommand \bibitemNoStop [0]{.\EOS\space}%
\providecommand \EOS [0]{\spacefactor3000\relax}%
\providecommand \BibitemShut  [1]{\csname bibitem#1\endcsname}%
\let\auto@bib@innerbib\@empty
\bibitem [{\citenamefont {Poggio}\ and\ \citenamefont
  {Anselmi}(2016)}]{poggio2016visual}%
  \BibitemOpen
  \bibfield  {author} {\bibinfo {author} {\bibfnamefont {T.~A.}\ \bibnamefont
  {Poggio}}\ and\ \bibinfo {author} {\bibfnamefont {F.}~\bibnamefont
  {Anselmi}},\ }\href@noop {} {\emph {\bibinfo {title} {Visual cortex and deep
  networks: learning invariant representations}}}\ (\bibinfo  {publisher} {MIT
  press},\ \bibinfo {year} {2016})\BibitemShut {NoStop}%
\bibitem [{\citenamefont {LeCun}\ \emph {et~al.}(2015)\citenamefont {LeCun},
  \citenamefont {Bengio},\ and\ \citenamefont {Hinton}}]{lecun2015deep}%
  \BibitemOpen
  \bibfield  {author} {\bibinfo {author} {\bibfnamefont {Y.}~\bibnamefont
  {LeCun}}, \bibinfo {author} {\bibfnamefont {Y.}~\bibnamefont {Bengio}},\ and\
  \bibinfo {author} {\bibfnamefont {G.}~\bibnamefont {Hinton}},\ }\href@noop {}
  {\bibfield  {journal} {\bibinfo  {journal} {nature}\ }\textbf {\bibinfo
  {volume} {521}},\ \bibinfo {pages} {436} (\bibinfo {year}
  {2015})}\BibitemShut {NoStop}%
\bibitem [{\citenamefont {Goodfellow}\ \emph {et~al.}(2009)\citenamefont
  {Goodfellow}, \citenamefont {Lee}, \citenamefont {Le}, \citenamefont {Saxe},\
  and\ \citenamefont {Ng}}]{goodfellow2009measuring}%
  \BibitemOpen
  \bibfield  {author} {\bibinfo {author} {\bibfnamefont {I.}~\bibnamefont
  {Goodfellow}}, \bibinfo {author} {\bibfnamefont {H.}~\bibnamefont {Lee}},
  \bibinfo {author} {\bibfnamefont {Q.}~\bibnamefont {Le}}, \bibinfo {author}
  {\bibfnamefont {A.}~\bibnamefont {Saxe}},\ and\ \bibinfo {author}
  {\bibfnamefont {A.}~\bibnamefont {Ng}},\ }\href@noop {} {\bibfield  {journal}
  {\bibinfo  {journal} {Advances in neural information processing systems}\
  }\textbf {\bibinfo {volume} {22}} (\bibinfo {year} {2009})}\BibitemShut
  {NoStop}%
\bibitem [{\citenamefont {Hong}\ \emph {et~al.}(2016)\citenamefont {Hong},
  \citenamefont {Yamins}, \citenamefont {Majaj},\ and\ \citenamefont
  {DiCarlo}}]{hong2016explicit}%
  \BibitemOpen
  \bibfield  {author} {\bibinfo {author} {\bibfnamefont {H.}~\bibnamefont
  {Hong}}, \bibinfo {author} {\bibfnamefont {D.~L.}\ \bibnamefont {Yamins}},
  \bibinfo {author} {\bibfnamefont {N.~J.}\ \bibnamefont {Majaj}},\ and\
  \bibinfo {author} {\bibfnamefont {J.~J.}\ \bibnamefont {DiCarlo}},\
  }\href@noop {} {\bibfield  {journal} {\bibinfo  {journal} {Nature
  neuroscience}\ }\textbf {\bibinfo {volume} {19}},\ \bibinfo {pages} {613}
  (\bibinfo {year} {2016})}\BibitemShut {NoStop}%
\bibitem [{\citenamefont {Drucker}\ \emph {et~al.}(1996)\citenamefont
  {Drucker}, \citenamefont {Burges}, \citenamefont {Kaufman}, \citenamefont
  {Smola},\ and\ \citenamefont {Vapnik}}]{vapnik1996svr}%
  \BibitemOpen
  \bibfield  {author} {\bibinfo {author} {\bibfnamefont {H.}~\bibnamefont
  {Drucker}}, \bibinfo {author} {\bibfnamefont {C.~J.~C.}\ \bibnamefont
  {Burges}}, \bibinfo {author} {\bibfnamefont {L.}~\bibnamefont {Kaufman}},
  \bibinfo {author} {\bibfnamefont {A.}~\bibnamefont {Smola}},\ and\ \bibinfo
  {author} {\bibfnamefont {V.}~\bibnamefont {Vapnik}},\ }in\ \href
  {https://proceedings.neurips.cc/paper/1996/file/d38901788c533e8286cb6400b40b386d-Paper.pdf}
  {\emph {\bibinfo {booktitle} {Advances in Neural Information Processing
  Systems}}},\ Vol.~\bibinfo {volume} {9},\ \bibinfo {editor} {edited by\
  \bibinfo {editor} {\bibfnamefont {M.}~\bibnamefont {Mozer}}, \bibinfo
  {editor} {\bibfnamefont {M.}~\bibnamefont {Jordan}},\ and\ \bibinfo {editor}
  {\bibfnamefont {T.}~\bibnamefont {Petsche}}}\ (\bibinfo  {publisher} {MIT
  Press},\ \bibinfo {year} {1996})\BibitemShut {NoStop}%
\bibitem [{\citenamefont {Sch{\"o}lkopf}\ \emph {et~al.}(2000)\citenamefont
  {Sch{\"o}lkopf}, \citenamefont {Smola}, \citenamefont {Williamson},\ and\
  \citenamefont {Bartlett}}]{scholkopf2000new}%
  \BibitemOpen
  \bibfield  {author} {\bibinfo {author} {\bibfnamefont {B.}~\bibnamefont
  {Sch{\"o}lkopf}}, \bibinfo {author} {\bibfnamefont {A.~J.}\ \bibnamefont
  {Smola}}, \bibinfo {author} {\bibfnamefont {R.~C.}\ \bibnamefont
  {Williamson}},\ and\ \bibinfo {author} {\bibfnamefont {P.~L.}\ \bibnamefont
  {Bartlett}},\ }\href@noop {} {\bibfield  {journal} {\bibinfo  {journal}
  {Neural computation}\ }\textbf {\bibinfo {volume} {12}},\ \bibinfo {pages}
  {1207} (\bibinfo {year} {2000})}\BibitemShut {NoStop}%
\bibitem [{\citenamefont {Belkin}\ \emph {et~al.}(2019)\citenamefont {Belkin},
  \citenamefont {Hsu}, \citenamefont {Ma},\ and\ \citenamefont
  {Mandal}}]{belkin2019reconciling}%
  \BibitemOpen
  \bibfield  {author} {\bibinfo {author} {\bibfnamefont {M.}~\bibnamefont
  {Belkin}}, \bibinfo {author} {\bibfnamefont {D.}~\bibnamefont {Hsu}},
  \bibinfo {author} {\bibfnamefont {S.}~\bibnamefont {Ma}},\ and\ \bibinfo
  {author} {\bibfnamefont {S.}~\bibnamefont {Mandal}},\ }\href@noop {}
  {\bibfield  {journal} {\bibinfo  {journal} {Proceedings of the National
  Academy of Sciences}\ }\textbf {\bibinfo {volume} {116}},\ \bibinfo {pages}
  {15849} (\bibinfo {year} {2019})}\BibitemShut {NoStop}%
\bibitem [{\citenamefont {M{\'e}zard}\ \emph {et~al.}(1987)\citenamefont
  {M{\'e}zard}, \citenamefont {Parisi},\ and\ \citenamefont
  {Virasoro}}]{mezard1987spin}%
  \BibitemOpen
  \bibfield  {author} {\bibinfo {author} {\bibfnamefont {M.}~\bibnamefont
  {M{\'e}zard}}, \bibinfo {author} {\bibfnamefont {G.}~\bibnamefont {Parisi}},\
  and\ \bibinfo {author} {\bibfnamefont {M.~A.}\ \bibnamefont {Virasoro}},\
  }\href@noop {} {\emph {\bibinfo {title} {Spin glass theory and beyond: An
  Introduction to the Replica Method and Its Applications}}},\ Vol.~\bibinfo
  {volume} {9}\ (\bibinfo  {publisher} {World Scientific Publishing Company},\
  \bibinfo {year} {1987})\BibitemShut {NoStop}%
\bibitem [{\citenamefont {Engel}(2001)}]{engel2001statistical}%
  \BibitemOpen
  \bibfield  {author} {\bibinfo {author} {\bibfnamefont {A.}~\bibnamefont
  {Engel}},\ }\href@noop {} {\emph {\bibinfo {title} {Statistical mechanics of
  learning}}}\ (\bibinfo  {publisher} {Cambridge University Press},\ \bibinfo
  {year} {2001})\BibitemShut {NoStop}%
\bibitem [{\citenamefont {Mezard}\ and\ \citenamefont
  {Montanari}(2009)}]{mezard2009information}%
  \BibitemOpen
  \bibfield  {author} {\bibinfo {author} {\bibfnamefont {M.}~\bibnamefont
  {Mezard}}\ and\ \bibinfo {author} {\bibfnamefont {A.}~\bibnamefont
  {Montanari}},\ }\href@noop {} {\emph {\bibinfo {title} {Information, physics,
  and computation}}}\ (\bibinfo  {publisher} {Oxford University Press},\
  \bibinfo {year} {2009})\BibitemShut {NoStop}%
\bibitem [{\citenamefont {Bordelon}\ \emph {et~al.}(2020)\citenamefont
  {Bordelon}, \citenamefont {Canatar},\ and\ \citenamefont
  {Pehlevan}}]{bordelon2020spectrum}%
  \BibitemOpen
  \bibfield  {author} {\bibinfo {author} {\bibfnamefont {B.}~\bibnamefont
  {Bordelon}}, \bibinfo {author} {\bibfnamefont {A.}~\bibnamefont {Canatar}},\
  and\ \bibinfo {author} {\bibfnamefont {C.}~\bibnamefont {Pehlevan}},\ }in\
  \href@noop {} {\emph {\bibinfo {booktitle} {International Conference on
  Machine Learning}}}\ (\bibinfo {organization} {PMLR},\ \bibinfo {year}
  {2020})\ pp.\ \bibinfo {pages} {1024--1034}\BibitemShut {NoStop}%
\bibitem [{\citenamefont {Jacot}\ \emph {et~al.}(2020)\citenamefont {Jacot},
  \citenamefont {Simsek}, \citenamefont {Spadaro}, \citenamefont {Hongler},\
  and\ \citenamefont {Gabriel}}]{jacot2020implicit}%
  \BibitemOpen
  \bibfield  {author} {\bibinfo {author} {\bibfnamefont {A.}~\bibnamefont
  {Jacot}}, \bibinfo {author} {\bibfnamefont {B.}~\bibnamefont {Simsek}},
  \bibinfo {author} {\bibfnamefont {F.}~\bibnamefont {Spadaro}}, \bibinfo
  {author} {\bibfnamefont {C.}~\bibnamefont {Hongler}},\ and\ \bibinfo {author}
  {\bibfnamefont {F.}~\bibnamefont {Gabriel}},\ }in\ \href@noop {} {\emph
  {\bibinfo {booktitle} {International Conference on Machine Learning}}}\
  (\bibinfo {organization} {PMLR},\ \bibinfo {year} {2020})\ pp.\ \bibinfo
  {pages} {4631--4640}\BibitemShut {NoStop}%
\bibitem [{\citenamefont {Canatar}\ \emph {et~al.}(2021)\citenamefont
  {Canatar}, \citenamefont {Bordelon},\ and\ \citenamefont
  {Pehlevan}}]{canatar2021spectral}%
  \BibitemOpen
  \bibfield  {author} {\bibinfo {author} {\bibfnamefont {A.}~\bibnamefont
  {Canatar}}, \bibinfo {author} {\bibfnamefont {B.}~\bibnamefont {Bordelon}},\
  and\ \bibinfo {author} {\bibfnamefont {C.}~\bibnamefont {Pehlevan}},\
  }\href@noop {} {\bibfield  {journal} {\bibinfo  {journal} {Nature
  communications}\ }\textbf {\bibinfo {volume} {12}},\ \bibinfo {pages} {2914}
  (\bibinfo {year} {2021})}\BibitemShut {NoStop}%
\bibitem [{\citenamefont {Simon}\ \emph {et~al.}(2023)\citenamefont {Simon},
  \citenamefont {Dickens}, \citenamefont {Karkada},\ and\ \citenamefont
  {Deweese}}]{simon2023eigenlearning}%
  \BibitemOpen
  \bibfield  {author} {\bibinfo {author} {\bibfnamefont {J.~B.}\ \bibnamefont
  {Simon}}, \bibinfo {author} {\bibfnamefont {M.}~\bibnamefont {Dickens}},
  \bibinfo {author} {\bibfnamefont {D.}~\bibnamefont {Karkada}},\ and\ \bibinfo
  {author} {\bibfnamefont {M.}~\bibnamefont {Deweese}},\ }\href@noop {}
  {\bibfield  {journal} {\bibinfo  {journal} {Transactions on Machine Learning
  Research}\ } (\bibinfo {year} {2023})}\BibitemShut {NoStop}%
\bibitem [{\citenamefont {Atanasov}\ \emph {et~al.}(2022)\citenamefont
  {Atanasov}, \citenamefont {Bordelon}, \citenamefont {Sainathan},\ and\
  \citenamefont {Pehlevan}}]{atanasov2022onset}%
  \BibitemOpen
  \bibfield  {author} {\bibinfo {author} {\bibfnamefont {A.}~\bibnamefont
  {Atanasov}}, \bibinfo {author} {\bibfnamefont {B.}~\bibnamefont {Bordelon}},
  \bibinfo {author} {\bibfnamefont {S.}~\bibnamefont {Sainathan}},\ and\
  \bibinfo {author} {\bibfnamefont {C.}~\bibnamefont {Pehlevan}},\ }in\
  \href@noop {} {\emph {\bibinfo {booktitle} {The Eleventh International
  Conference on Learning Representations}}}\ (\bibinfo {year}
  {2022})\BibitemShut {NoStop}%
\bibitem [{\citenamefont {Atanasov}\ \emph {et~al.}(2024)\citenamefont
  {Atanasov}, \citenamefont {Zavatone-Veth},\ and\ \citenamefont
  {Pehlevan}}]{atanasov2024scaling}%
  \BibitemOpen
  \bibfield  {author} {\bibinfo {author} {\bibfnamefont {A.~B.}\ \bibnamefont
  {Atanasov}}, \bibinfo {author} {\bibfnamefont {J.~A.}\ \bibnamefont
  {Zavatone-Veth}},\ and\ \bibinfo {author} {\bibfnamefont {C.}~\bibnamefont
  {Pehlevan}},\ }\href@noop {} {\bibinfo {title} {Scaling and renormalization
  in high-dimensional regression}} (\bibinfo {year} {2024}),\ \Eprint
  {https://arxiv.org/abs/2405.00592} {arXiv:2405.00592 [stat.ML]} \BibitemShut
  {NoStop}%
\bibitem [{\citenamefont {Loureiro}\ \emph {et~al.}(2021)\citenamefont
  {Loureiro}, \citenamefont {Gerbelot}, \citenamefont {Cui}, \citenamefont
  {Goldt}, \citenamefont {Krzakala}, \citenamefont {Mezard},\ and\
  \citenamefont {Zdeborov{\'a}}}]{loureiro2021learning}%
  \BibitemOpen
  \bibfield  {author} {\bibinfo {author} {\bibfnamefont {B.}~\bibnamefont
  {Loureiro}}, \bibinfo {author} {\bibfnamefont {C.}~\bibnamefont {Gerbelot}},
  \bibinfo {author} {\bibfnamefont {H.}~\bibnamefont {Cui}}, \bibinfo {author}
  {\bibfnamefont {S.}~\bibnamefont {Goldt}}, \bibinfo {author} {\bibfnamefont
  {F.}~\bibnamefont {Krzakala}}, \bibinfo {author} {\bibfnamefont
  {M.}~\bibnamefont {Mezard}},\ and\ \bibinfo {author} {\bibfnamefont
  {L.}~\bibnamefont {Zdeborov{\'a}}},\ }\href@noop {} {\bibfield  {journal}
  {\bibinfo  {journal} {Advances in Neural Information Processing Systems}\
  }\textbf {\bibinfo {volume} {34}},\ \bibinfo {pages} {18137} (\bibinfo {year}
  {2021})}\BibitemShut {NoStop}%
\bibitem [{\citenamefont {Bengio}\ \emph {et~al.}(2013)\citenamefont {Bengio},
  \citenamefont {Courville},\ and\ \citenamefont
  {Vincent}}]{bengio2013representation}%
  \BibitemOpen
  \bibfield  {author} {\bibinfo {author} {\bibfnamefont {Y.}~\bibnamefont
  {Bengio}}, \bibinfo {author} {\bibfnamefont {A.}~\bibnamefont {Courville}},\
  and\ \bibinfo {author} {\bibfnamefont {P.}~\bibnamefont {Vincent}},\
  }\href@noop {} {\bibfield  {journal} {\bibinfo  {journal} {IEEE transactions
  on pattern analysis and machine intelligence}\ }\textbf {\bibinfo {volume}
  {35}},\ \bibinfo {pages} {1798} (\bibinfo {year} {2013})}\BibitemShut
  {NoStop}%
\bibitem [{\citenamefont {Achille}\ and\ \citenamefont
  {Soatto}(2018)}]{achille2018emergence}%
  \BibitemOpen
  \bibfield  {author} {\bibinfo {author} {\bibfnamefont {A.}~\bibnamefont
  {Achille}}\ and\ \bibinfo {author} {\bibfnamefont {S.}~\bibnamefont
  {Soatto}},\ }\href@noop {} {\bibfield  {journal} {\bibinfo  {journal}
  {Journal of Machine Learning Research}\ }\textbf {\bibinfo {volume} {19}},\
  \bibinfo {pages} {1} (\bibinfo {year} {2018})}\BibitemShut {NoStop}%
\bibitem [{\citenamefont {Chung}\ and\ \citenamefont
  {Abbott}(2021)}]{chung2021neural}%
  \BibitemOpen
  \bibfield  {author} {\bibinfo {author} {\bibfnamefont {S.}~\bibnamefont
  {Chung}}\ and\ \bibinfo {author} {\bibfnamefont {L.}~\bibnamefont {Abbott}},\
  }\href@noop {} {\bibfield  {journal} {\bibinfo  {journal} {Curr. opin.
  neurobiol.}\ }\textbf {\bibinfo {volume} {70}},\ \bibinfo {pages} {137}
  (\bibinfo {year} {2021})}\BibitemShut {NoStop}%
\bibitem [{\citenamefont {Mei}\ \emph {et~al.}(2021)\citenamefont {Mei},
  \citenamefont {Misiakiewicz},\ and\ \citenamefont
  {Montanari}}]{mei2021learning}%
  \BibitemOpen
  \bibfield  {author} {\bibinfo {author} {\bibfnamefont {S.}~\bibnamefont
  {Mei}}, \bibinfo {author} {\bibfnamefont {T.}~\bibnamefont {Misiakiewicz}},\
  and\ \bibinfo {author} {\bibfnamefont {A.}~\bibnamefont {Montanari}},\ }in\
  \href@noop {} {\emph {\bibinfo {booktitle} {Conference on Learning Theory}}}\
  (\bibinfo {organization} {PMLR},\ \bibinfo {year} {2021})\ pp.\ \bibinfo
  {pages} {3351--3418}\BibitemShut {NoStop}%
\bibitem [{\citenamefont {Favero}\ \emph {et~al.}(2021)\citenamefont {Favero},
  \citenamefont {Cagnetta},\ and\ \citenamefont {Wyart}}]{favero2021locality}%
  \BibitemOpen
  \bibfield  {author} {\bibinfo {author} {\bibfnamefont {A.}~\bibnamefont
  {Favero}}, \bibinfo {author} {\bibfnamefont {F.}~\bibnamefont {Cagnetta}},\
  and\ \bibinfo {author} {\bibfnamefont {M.}~\bibnamefont {Wyart}},\
  }\href@noop {} {\bibfield  {journal} {\bibinfo  {journal} {Advances in Neural
  Information Processing Systems}\ }\textbf {\bibinfo {volume} {34}},\ \bibinfo
  {pages} {9456} (\bibinfo {year} {2021})}\BibitemShut {NoStop}%
\bibitem [{\citenamefont {Elesedy}(2021)}]{elesedy2021provably}%
  \BibitemOpen
  \bibfield  {author} {\bibinfo {author} {\bibfnamefont {B.}~\bibnamefont
  {Elesedy}},\ }\href@noop {} {\bibfield  {journal} {\bibinfo  {journal}
  {Advances in Neural Information Processing Systems}\ }\textbf {\bibinfo
  {volume} {34}},\ \bibinfo {pages} {17273} (\bibinfo {year}
  {2021})}\BibitemShut {NoStop}%
\bibitem [{\citenamefont {Chung}\ \emph {et~al.}(2018)\citenamefont {Chung},
  \citenamefont {Lee},\ and\ \citenamefont
  {Sompolinsky}}]{chung2018classification}%
  \BibitemOpen
  \bibfield  {author} {\bibinfo {author} {\bibfnamefont {S.}~\bibnamefont
  {Chung}}, \bibinfo {author} {\bibfnamefont {D.~D.}\ \bibnamefont {Lee}},\
  and\ \bibinfo {author} {\bibfnamefont {H.}~\bibnamefont {Sompolinsky}},\
  }\href@noop {} {\bibfield  {journal} {\bibinfo  {journal} {Physical Review
  X}\ }\textbf {\bibinfo {volume} {8}},\ \bibinfo {pages} {031003} (\bibinfo
  {year} {2018})}\BibitemShut {NoStop}%
\bibitem [{\citenamefont {Wakhloo}\ \emph {et~al.}(2023)\citenamefont
  {Wakhloo}, \citenamefont {Sussman},\ and\ \citenamefont
  {Chung}}]{wakhloo2023linear}%
  \BibitemOpen
  \bibfield  {author} {\bibinfo {author} {\bibfnamefont {A.~J.}\ \bibnamefont
  {Wakhloo}}, \bibinfo {author} {\bibfnamefont {T.~J.}\ \bibnamefont
  {Sussman}},\ and\ \bibinfo {author} {\bibfnamefont {S.}~\bibnamefont
  {Chung}},\ }\href@noop {} {\bibfield  {journal} {\bibinfo  {journal}
  {Physical Review Letters}\ }\textbf {\bibinfo {volume} {131}},\ \bibinfo
  {pages} {027301} (\bibinfo {year} {2023})}\BibitemShut {NoStop}%
\bibitem [{\citenamefont {Farrell}\ \emph {et~al.}(2021)\citenamefont
  {Farrell}, \citenamefont {Bordelon}, \citenamefont {Trivedi},\ and\
  \citenamefont {Pehlevan}}]{farrell2021capacity}%
  \BibitemOpen
  \bibfield  {author} {\bibinfo {author} {\bibfnamefont {M.}~\bibnamefont
  {Farrell}}, \bibinfo {author} {\bibfnamefont {B.}~\bibnamefont {Bordelon}},
  \bibinfo {author} {\bibfnamefont {S.}~\bibnamefont {Trivedi}},\ and\ \bibinfo
  {author} {\bibfnamefont {C.}~\bibnamefont {Pehlevan}},\ }\href@noop {}
  {\bibfield  {journal} {\bibinfo  {journal} {arXiv preprint arXiv:2110.07472}\
  } (\bibinfo {year} {2021})}\BibitemShut {NoStop}%
\bibitem [{\citenamefont {Ruben}\ and\ \citenamefont
  {Pehlevan}(2023)}]{ruben2023learning}%
  \BibitemOpen
  \bibfield  {author} {\bibinfo {author} {\bibfnamefont {B.}~\bibnamefont
  {Ruben}}\ and\ \bibinfo {author} {\bibfnamefont {C.}~\bibnamefont
  {Pehlevan}},\ }\href@noop {} {\bibfield  {journal} {\bibinfo  {journal}
  {Advances in Neural Information Processing Systems}\ }\textbf {\bibinfo
  {volume} {36}},\ \bibinfo {pages} {50041} (\bibinfo {year}
  {2023})}\BibitemShut {NoStop}%
\bibitem [{\citenamefont {Adlam}\ and\ \citenamefont
  {Pennington}(2020)}]{adlam2020neural}%
  \BibitemOpen
  \bibfield  {author} {\bibinfo {author} {\bibfnamefont {B.}~\bibnamefont
  {Adlam}}\ and\ \bibinfo {author} {\bibfnamefont {J.}~\bibnamefont
  {Pennington}},\ }in\ \href@noop {} {\emph {\bibinfo {booktitle}
  {International Conference on Machine Learning}}}\ (\bibinfo {organization}
  {PMLR},\ \bibinfo {year} {2020})\ pp.\ \bibinfo {pages} {74--84}\BibitemShut
  {NoStop}%
\bibitem [{\citenamefont {Cheng}\ and\ \citenamefont
  {Montanari}(2022)}]{cheng2022dimension}%
  \BibitemOpen
  \bibfield  {author} {\bibinfo {author} {\bibfnamefont {C.}~\bibnamefont
  {Cheng}}\ and\ \bibinfo {author} {\bibfnamefont {A.}~\bibnamefont
  {Montanari}},\ }\href@noop {} {\bibinfo {title} {Dimension free ridge
  regression}} (\bibinfo {year} {2022}),\ \Eprint
  {https://arxiv.org/abs/2210.08571} {arXiv:2210.08571 [math.ST]} \BibitemShut
  {NoStop}%
\bibitem [{\citenamefont {B{\"o}s}\ \emph {et~al.}(1993)\citenamefont
  {B{\"o}s}, \citenamefont {Kinzel},\ and\ \citenamefont
  {Opper}}]{bos1993generalization}%
  \BibitemOpen
  \bibfield  {author} {\bibinfo {author} {\bibfnamefont {S.}~\bibnamefont
  {B{\"o}s}}, \bibinfo {author} {\bibfnamefont {W.}~\bibnamefont {Kinzel}},\
  and\ \bibinfo {author} {\bibfnamefont {M.}~\bibnamefont {Opper}},\
  }\href@noop {} {\bibfield  {journal} {\bibinfo  {journal} {Physical Review
  E}\ }\textbf {\bibinfo {volume} {47}},\ \bibinfo {pages} {1384} (\bibinfo
  {year} {1993})}\BibitemShut {NoStop}%
\bibitem [{\citenamefont {Seung}\ and\ \citenamefont
  {Sompolinsky}(1993)}]{seung1993simple}%
  \BibitemOpen
  \bibfield  {author} {\bibinfo {author} {\bibfnamefont {H.~S.}\ \bibnamefont
  {Seung}}\ and\ \bibinfo {author} {\bibfnamefont {H.}~\bibnamefont
  {Sompolinsky}},\ }\href@noop {} {\bibfield  {journal} {\bibinfo  {journal}
  {Proceedings of the national academy of sciences}\ }\textbf {\bibinfo
  {volume} {90}},\ \bibinfo {pages} {10749} (\bibinfo {year}
  {1993})}\BibitemShut {NoStop}%
\bibitem [{\citenamefont {Brunel}\ and\ \citenamefont
  {Nadal}(1998)}]{brunel1998mutual}%
  \BibitemOpen
  \bibfield  {author} {\bibinfo {author} {\bibfnamefont {N.}~\bibnamefont
  {Brunel}}\ and\ \bibinfo {author} {\bibfnamefont {J.-P.}\ \bibnamefont
  {Nadal}},\ }\href@noop {} {\bibfield  {journal} {\bibinfo  {journal} {Neural
  computation}\ }\textbf {\bibinfo {volume} {10}},\ \bibinfo {pages} {1731}
  (\bibinfo {year} {1998})}\BibitemShut {NoStop}%
\bibitem [{\citenamefont {Belkin}\ \emph {et~al.}(2018)\citenamefont {Belkin},
  \citenamefont {Ma},\ and\ \citenamefont {Mandal}}]{belkin2018understand}%
  \BibitemOpen
  \bibfield  {author} {\bibinfo {author} {\bibfnamefont {M.}~\bibnamefont
  {Belkin}}, \bibinfo {author} {\bibfnamefont {S.}~\bibnamefont {Ma}},\ and\
  \bibinfo {author} {\bibfnamefont {S.}~\bibnamefont {Mandal}},\ }in\
  \href@noop {} {\emph {\bibinfo {booktitle} {International Conference on
  Machine Learning}}}\ (\bibinfo {organization} {PMLR},\ \bibinfo {year}
  {2018})\ pp.\ \bibinfo {pages} {541--549}\BibitemShut {NoStop}%
\bibitem [{\citenamefont {Sch{\"o}lkopf}\ \emph {et~al.}(2002)\citenamefont
  {Sch{\"o}lkopf}, \citenamefont {Smola}, \citenamefont {Bach} \emph
  {et~al.}}]{scholkopf2002learning}%
  \BibitemOpen
  \bibfield  {author} {\bibinfo {author} {\bibfnamefont {B.}~\bibnamefont
  {Sch{\"o}lkopf}}, \bibinfo {author} {\bibfnamefont {A.~J.}\ \bibnamefont
  {Smola}}, \bibinfo {author} {\bibfnamefont {F.}~\bibnamefont {Bach}}, \emph
  {et~al.},\ }\href@noop {} {\emph {\bibinfo {title} {Learning with kernels:
  support vector machines, regularization, optimization, and beyond}}}\
  (\bibinfo  {publisher} {MIT press},\ \bibinfo {year} {2002})\BibitemShut
  {NoStop}%
\bibitem [{\citenamefont {Maloney}\ \emph {et~al.}(2022)\citenamefont
  {Maloney}, \citenamefont {Roberts},\ and\ \citenamefont
  {Sully}}]{maloney2022solvable}%
  \BibitemOpen
  \bibfield  {author} {\bibinfo {author} {\bibfnamefont {A.}~\bibnamefont
  {Maloney}}, \bibinfo {author} {\bibfnamefont {D.~A.}\ \bibnamefont
  {Roberts}},\ and\ \bibinfo {author} {\bibfnamefont {J.}~\bibnamefont
  {Sully}},\ }\href@noop {} {\bibfield  {journal} {\bibinfo  {journal} {arXiv
  preprint arXiv:2210.16859}\ } (\bibinfo {year} {2022})}\BibitemShut {NoStop}%
\bibitem [{\citenamefont {Wu}\ and\ \citenamefont {Xu}(2020)}]{wu2020optimal}%
  \BibitemOpen
  \bibfield  {author} {\bibinfo {author} {\bibfnamefont {D.}~\bibnamefont
  {Wu}}\ and\ \bibinfo {author} {\bibfnamefont {J.}~\bibnamefont {Xu}},\
  }\href@noop {} {\bibfield  {journal} {\bibinfo  {journal} {Advances in Neural
  Information Processing Systems}\ }\textbf {\bibinfo {volume} {33}},\ \bibinfo
  {pages} {10112} (\bibinfo {year} {2020})}\BibitemShut {NoStop}%
\bibitem [{\citenamefont {Hastie}(2009)}]{hastie2009elements}%
  \BibitemOpen
  \bibfield  {author} {\bibinfo {author} {\bibfnamefont {T.}~\bibnamefont
  {Hastie}},\ }\href@noop {} {\bibinfo {title} {The elements of statistical
  learning: data mining, inference, and prediction}} (\bibinfo {year}
  {2009})\BibitemShut {NoStop}%
\bibitem [{\citenamefont {Canatar}\ \emph {et~al.}(2024)\citenamefont
  {Canatar}, \citenamefont {Feather}, \citenamefont {Wakhloo},\ and\
  \citenamefont {Chung}}]{canatar2024spectral}%
  \BibitemOpen
  \bibfield  {author} {\bibinfo {author} {\bibfnamefont {A.}~\bibnamefont
  {Canatar}}, \bibinfo {author} {\bibfnamefont {J.}~\bibnamefont {Feather}},
  \bibinfo {author} {\bibfnamefont {A.}~\bibnamefont {Wakhloo}},\ and\ \bibinfo
  {author} {\bibfnamefont {S.}~\bibnamefont {Chung}},\ }\href@noop {}
  {\bibfield  {journal} {\bibinfo  {journal} {Advances in Neural Information
  Processing Systems}\ }\textbf {\bibinfo {volume} {36}} (\bibinfo {year}
  {2024})}\BibitemShut {NoStop}%
\bibitem [{\citenamefont {Thrampoulidis}\ \emph {et~al.}(2015)\citenamefont
  {Thrampoulidis}, \citenamefont {Oymak},\ and\ \citenamefont
  {Hassibi}}]{thrampoulidis2015regularized}%
  \BibitemOpen
  \bibfield  {author} {\bibinfo {author} {\bibfnamefont {C.}~\bibnamefont
  {Thrampoulidis}}, \bibinfo {author} {\bibfnamefont {S.}~\bibnamefont
  {Oymak}},\ and\ \bibinfo {author} {\bibfnamefont {B.}~\bibnamefont
  {Hassibi}},\ }in\ \href@noop {} {\emph {\bibinfo {booktitle} {Conference on
  Learning Theory}}}\ (\bibinfo {organization} {PMLR},\ \bibinfo {year}
  {2015})\ pp.\ \bibinfo {pages} {1683--1709}\BibitemShut {NoStop}%
\bibitem [{\citenamefont {Dobriban}\ and\ \citenamefont
  {Wager}(2018)}]{dobriban2018high}%
  \BibitemOpen
  \bibfield  {author} {\bibinfo {author} {\bibfnamefont {E.}~\bibnamefont
  {Dobriban}}\ and\ \bibinfo {author} {\bibfnamefont {S.}~\bibnamefont
  {Wager}},\ }\href@noop {} {\bibfield  {journal} {\bibinfo  {journal} {The
  Annals of Statistics}\ }\textbf {\bibinfo {volume} {46}},\ \bibinfo {pages}
  {247} (\bibinfo {year} {2018})}\BibitemShut {NoStop}%
\bibitem [{\citenamefont {Kobak}\ \emph {et~al.}(2020)\citenamefont {Kobak},
  \citenamefont {Lomond},\ and\ \citenamefont {Sanchez}}]{kobak2020optimal}%
  \BibitemOpen
  \bibfield  {author} {\bibinfo {author} {\bibfnamefont {D.}~\bibnamefont
  {Kobak}}, \bibinfo {author} {\bibfnamefont {J.}~\bibnamefont {Lomond}},\ and\
  \bibinfo {author} {\bibfnamefont {B.}~\bibnamefont {Sanchez}},\ }\href@noop
  {} {\bibfield  {journal} {\bibinfo  {journal} {The Journal of Machine
  Learning Research}\ }\textbf {\bibinfo {volume} {21}},\ \bibinfo {pages}
  {6863} (\bibinfo {year} {2020})}\BibitemShut {NoStop}%
\bibitem [{\citenamefont {Hastie}\ \emph {et~al.}(2022)\citenamefont {Hastie},
  \citenamefont {Montanari}, \citenamefont {Rosset},\ and\ \citenamefont
  {Tibshirani}}]{hastie2022surprises}%
  \BibitemOpen
  \bibfield  {author} {\bibinfo {author} {\bibfnamefont {T.}~\bibnamefont
  {Hastie}}, \bibinfo {author} {\bibfnamefont {A.}~\bibnamefont {Montanari}},
  \bibinfo {author} {\bibfnamefont {S.}~\bibnamefont {Rosset}},\ and\ \bibinfo
  {author} {\bibfnamefont {R.~J.}\ \bibnamefont {Tibshirani}},\ }\href@noop {}
  {\bibfield  {journal} {\bibinfo  {journal} {The Annals of Statistics}\
  }\textbf {\bibinfo {volume} {50}},\ \bibinfo {pages} {949} (\bibinfo {year}
  {2022})}\BibitemShut {NoStop}%
\bibitem [{\citenamefont {Zhang}(2024)}]{zhang2024neural}%
  \BibitemOpen
  \bibfield  {author} {\bibinfo {author} {\bibfnamefont {Z.}~\bibnamefont
  {Zhang}},\ }\href@noop {} {\bibfield  {journal} {\bibinfo  {journal} {arXiv
  preprint arXiv:2405.19398}\ } (\bibinfo {year} {2024})}\BibitemShut {NoStop}%
\bibitem [{\citenamefont {Mei}\ \emph {et~al.}(2022)\citenamefont {Mei},
  \citenamefont {Misiakiewicz},\ and\ \citenamefont
  {Montanari}}]{mei2022generalization}%
  \BibitemOpen
  \bibfield  {author} {\bibinfo {author} {\bibfnamefont {S.}~\bibnamefont
  {Mei}}, \bibinfo {author} {\bibfnamefont {T.}~\bibnamefont {Misiakiewicz}},\
  and\ \bibinfo {author} {\bibfnamefont {A.}~\bibnamefont {Montanari}},\
  }\href@noop {} {\bibfield  {journal} {\bibinfo  {journal} {Applied and
  Computational Harmonic Analysis}\ }\textbf {\bibinfo {volume} {59}},\
  \bibinfo {pages} {3} (\bibinfo {year} {2022})}\BibitemShut {NoStop}%
\bibitem [{\citenamefont {Mei}\ and\ \citenamefont
  {Montanari}(2022)}]{mei2022generalization2}%
  \BibitemOpen
  \bibfield  {author} {\bibinfo {author} {\bibfnamefont {S.}~\bibnamefont
  {Mei}}\ and\ \bibinfo {author} {\bibfnamefont {A.}~\bibnamefont
  {Montanari}},\ }\href@noop {} {\bibfield  {journal} {\bibinfo  {journal}
  {Communications on Pure and Applied Mathematics}\ }\textbf {\bibinfo {volume}
  {75}},\ \bibinfo {pages} {667} (\bibinfo {year} {2022})}\BibitemShut
  {NoStop}%
\bibitem [{\citenamefont {Hu}\ and\ \citenamefont {Lu}(2022)}]{hu2022sharp}%
  \BibitemOpen
  \bibfield  {author} {\bibinfo {author} {\bibfnamefont {H.}~\bibnamefont
  {Hu}}\ and\ \bibinfo {author} {\bibfnamefont {Y.~M.}\ \bibnamefont {Lu}},\
  }\href@noop {} {\bibfield  {journal} {\bibinfo  {journal} {arXiv preprint
  arXiv:2205.06798}\ } (\bibinfo {year} {2022})}\BibitemShut {NoStop}%
\bibitem [{\citenamefont {Misiakiewicz}\ and\ \citenamefont
  {Saeed}(2024)}]{misiakiewicz2024nonasymptotic}%
  \BibitemOpen
  \bibfield  {author} {\bibinfo {author} {\bibfnamefont {T.}~\bibnamefont
  {Misiakiewicz}}\ and\ \bibinfo {author} {\bibfnamefont {B.}~\bibnamefont
  {Saeed}},\ }\href@noop {} {\bibinfo {title} {A non-asymptotic theory of
  kernel ridge regression: deterministic equivalents, test error, and gcv
  estimator}} (\bibinfo {year} {2024}),\ \Eprint
  {https://arxiv.org/abs/2403.08938} {arXiv:2403.08938 [stat.ML]} \BibitemShut
  {NoStop}%
\end{thebibliography}%

\setcounter{figure}{0}
\renewcommand{\thefigure}{S\arabic{figure}}
\renewcommand{\theHfigure}{S\arabic{figure}}
\appendix
\onecolumngrid

\section{Problem Setup}\label{sec:SI_replica}

Here, we study a family of regression problems and their generalization properties using the replica method. In our notation, we consider a training set $\cD = \{(\x^\mu, \bar y^\mu)\}_{\mu=1}^P$ consisting of $P$ i.i.d. drawn inputs $\x^\mu$ and labels $\bar y^\mu$. In this work, we assume that the target labels are related to the inputs via
\begin{align}
    \bar y^\mu = \bar\w\cdot\bar\bpsi^\mu, \quad \bar\bpsi^\mu \equiv \bar \bpsi(\x^\mu) \in \bR^N, \quad \bar\w \in \bR^N,
\end{align}
where $\bar \bpsi^\mu$ denote $N$-dimensional \textit{center} features as a function of input samples, and $\bar\w$ denotes the \textit{coding direction}. On the other hand, we consider a linear predictor 
\begin{align}
    y^\mu = \w\cdot\bpsi^\mu, \quad \bpsi^\mu \equiv \bpsi(\x^\mu) \in \bR^N, \quad \w \in \bR^N,
\end{align}
where $\bpsi^\mu$ denote the $N$-dimensional \textit{encoding} features available to the predictor, and $\w$ is to be learned.

Our generic regression task solves the following problem:
\begin{align}
    \min_\w \frac{1}{2} \norm{\w}^2 + \frac{1}{\lambda}\sum_{\mu=1}^P \ell\lrpar{\w \cdot \bpsi^\mu - \bar\w\cdot\bar\bpsi^\mu},
\end{align}
where $\ell(x)$ is the loss function. Below we list possible choices for $\ell(x)$:
\begin{itemize}
    \item \textbf{Kernel Ridge Regression (KRR)}: For ridge regression, $\ell(x) = \frac{1}{2}x^2$ and the optimization problem becomes:
    \begin{align}
        \min_\w \frac{1}{2} \norm{\w}^2 + \frac{1}{2\lambda}\sum_{\mu=1}^P \lrpar{\w \cdot \bpsi^\mu - \bar\w\cdot\bar\bpsi^\mu}^2.
    \end{align}
    Here, $\lambda\to 0$ limit corresponds to the ridgeless (least-squares) regression, and the case with linear features ($\bpsi(\x) = \x$) corresponds to linear regression.

    \item \textbf{Support Vector Regression (SVR)}: Specifically, we consider $\varepsilon-$insensitive SVR where the loss vanishes if the error is less than some positive tolerance variable $\varepsilon$. Generically, the loss function is $\ell(x) = \frac{1}{n!}\max(0, \abs{x} - \varepsilon)^n$, and the optimization problem becomes:
    \begin{align}
        \min_\w \frac{1}{2} \norm{\w}^2 + \frac{1}{n!\lambda}\sum_{\mu=1}^P \max\lrpar{0, \abs{\w \cdot \bpsi^\mu - \bar\w\cdot\bar\bpsi^\mu} - \varepsilon}^n,
    \end{align}
    where $n \in \bN_0$ are non-negative integers. Note that the case where $n=2$ and $\varepsilon=0$ is identical to KRR above.
\end{itemize}

A few notes are in order: 
\begin{itemize}

    \item Our calculations require averaging over this inputs $\x^\mu$, which is intractable for general feature maps. However, we make a \textit{Gaussian assumption} that the statistics of $\bar\bpsi(\x^\mu)$ and $\bpsi(\x^\mu)$ can be approximated by their second moments:
    \begin{align}\label{eq:SI_gaussian_assumption}
    \braket{\bar\bpsi(\x)\bar\bpsi(\x)^\top}_\x =  \frac{1}{N}\bSigma_{\bar\bpsi}, \quad \braket{\bpsi(\x)\bpsi(\x)^\top}_\x = \frac{1}{N}\bSigma_{\bpsi}, \quad \braket{\bpsi(\x)\bar\bpsi(\x)^\top}_\x = \frac{1}{N}\bSigma_{\bpsi\bar\bpsi}
    \end{align}
    and the final results will depend only on these quantities. This approximation simplifies our calculations, yields excellent agreement with experiments, and is also justified by the \textit{Gaussian equivalence} theorems in high dimensional statistics where the covariance of non-linear features (kernels) can be summarized by an equivalent Gaussian statistics \cite{mei2022generalization, mei2022generalization2, hu2022sharp, misiakiewicz2024nonasymptotic}. 

    \item In our formulation, we excluded the effect of label noise in the training set \cite{canatar2021spectral} to avoid clutter, since the effect of label noise with variance $\sigma^2$ can equivalently be studied by shifting $\bar\bpsi^\mu \to \bar\bpsi^\mu + \n^\mu$ with an independent random vector $\n$ with $\bE (\bar\w\cdot\n)^2 = \sigma^2$.

    \item We assume that the encoding representations depend on centers via
    \begin{align}
        \bpsi(\x) = \A\bar\bpsi(\x) + \bdelta(\x)
    \end{align}
    Here, $\A \in \bR^{M\times N}$ is a deterministic projection matrix that models the relation between centers and encoding features. The noise representation $\bdelta(\x) \in \bR^M$ models the \textit{variations} in the encoding features, and is the source of randomness in $\bpsi(\x)$ for each fixed input $\x$.
    
    \item While, in general, $\bdelta(\x)$ should be thought of as a random feature conditioned on each input $\x$, here we consider a noise model independent of centers. The simplest model for $\bdelta(\x)$ with input correlations is a linear random feature model $\bdelta(\x) = \B \bar\bpsi(\x) + \bdelta_0$ where $\B \in \bR^{M\times N}$ is a random matrix, and $\bdelta_0$ is a random vector which are drawn independently from their respective distributions. The study of this model requires computing another quenched average over random features $\B$ and is left for future work. See \cite{atanasov2022onset, maloney2022solvable} for the analysis of a similar model in ridge regression setting.

    \item In particular, we will apply our theory to the case where $\bpsi(\x) = \bar\bpsi(\x) + \bdelta$ where $\bdelta \in \bR^N$ models the variances in the inputs that are irrelevant to the target. The dimensionalities of the features $\bpsi$ and $\bar\bpsi$ are chosen to be the same for notational convenience, however, can be trivially applied to a more general setting with $\bpsi = \A\bar\bpsi + \bdelta$, since our final result will depend only on the covariance matrices in Eq.\eqref{eq:SI_gaussian_assumption} and the coding direction $\bar\w$.
\end{itemize}

\subsection{Replica Calculation}

We set up a general framework for our calculation. The measure for the weights are denoted as $d\mu(\w)$, and the loss function $\ell\lrpar{\w\cdot\bpsi - \bar\w\cdot\bar\bpsi}$ is kept generic. Then, the partition function is given by
\begin{align}
     Z = \int d\tilde\mu(\w)e^{- \frac{\beta}{\lambda} \sum_{\mu=1}^P \ell\lrpar{\w \cdot\bpsi^\mu - \bar\w\cdot \bar\bpsi^{\mu}}},\quad d\tilde\mu(\w) \equiv d\mu(\w)e^{-\beta \w^\top \J_{\bpsi\bar\bpsi}\bar\w}
\end{align}
where we defined $d\tilde\mu(\w)$ by including a source term to extract the mean and variance of the optimized weights. 

First, we integrate in auxiliary variables to simplify the computation:
\begin{align}
    H_{\mu} \equiv \w\cdot\bpsi^\mu - \bar\w\cdot \bar\bpsi^\mu 
\end{align}
and get:
\begin{align}
    Z &= \int d\tilde\mu(\w) \prod_{\mu} dH_{\mu} \, e^{- \frac{\beta}{\lambda} \ell(H_{\mu})} \delta\lrpar{H_{\mu} - \w\cdot\bpsi^\mu + \bar\w\cdot \bar\bpsi^\mu}\nonumber\\
    &= \int d\tilde\mu(\w) [d\cH_{\mu}] [d\hat H_{\mu}] \,e^{i\sum_{\mu}\hat H_{\mu} H_{\mu}} e^{-i\sum_{\mu}\hat H_{\mu} \lrpar{\w\cdot\bpsi^\mu - \bar\w\cdot \bar\bpsi^\mu  }},
\end{align}
where we defined $d\cH_{\mu} \equiv dH_{\mu}\, e^{- \frac{\beta}{\lambda} \ell(H_{\mu})}$ and introduced the convention where $[dx^{a,b,\dots}]$ represents product over free indices $[dx^{a,b,\dots}] \equiv \prod_{a,b,\dots} dx^{a,b,\dots}$. For example, $[d\cH_{\mu}] \equiv \prod_{\mu} d\cH_{\mu}$ and $[d\hat H_{\mu}] \equiv \prod_{\mu} d\hat H_{\mu}$.

Next, we replicate the partition function to perform averages over the data distribution:
\begin{align}
    Z^n &= \int [d\tilde\mu(\w^a)] [d\cH^{a}_{\mu}] [d\hat H^{a}_{\mu}] \,e^{i\sum_{a,\mu}\hat H^{a}_{\mu} H^{a}_{\mu}} e^{-i\sum_{a,\mu}\hat H^{a}_{\mu} \lrpar{\w^a\cdot\bpsi^\mu - \bar\w\cdot \bar\bpsi^{\mu}}}.
\end{align}
Next, we define the $(2N)$-dimensional feature vector $\bgamma^{\mu}$ and $(2N)$-dimensional weight vector $\bm{\cW}^a$ as
\begin{align}
    \bgamma^{\mu} = \begin{pmatrix}
        \bpsi^\mu \\ \bar\bpsi^{\mu}
    \end{pmatrix}, \quad \bm{\cW}^a = \begin{pmatrix}
        \w^a \\ -\bar\w
    \end{pmatrix},
\end{align}
so that for each sample, we have:
\begin{align}
    \w^a\cdot\bpsi^\mu - \bar\w\cdot \bar\bpsi^{\mu} = \bm{\cW}^a\cdot \bgamma^{\mu}.
\end{align}
At this point, we make the Gaussian assumption where the statistics of $\bgamma^\mu$ can be captured by its first and second moments:
\begin{align}
    \braket{\bgamma} = 0, \quad \braket{\bgamma\bgamma^\top} = \begin{pmatrix}
        \braket{\bpsi\bpsi^\top} & \braket{\bpsi\bar\bpsi^\top}\\
        \braket{\bar\bpsi \bpsi^\top} & \braket{\bar\bpsi\bar\bpsi^\top}
    \end{pmatrix} = \begin{pmatrix}
        \bSigma_\bpsi / N & \bSigma_{\bpsi\bar\bpsi} / N\\
        \bSigma_{\bpsi\bar\bpsi}^\top / N & \bSigma_{\bar\bpsi} / N
    \end{pmatrix} \equiv \bSigma_\bGamma
\end{align}
where $\bSigma_\bGamma \in \bR^{(2N) \times (2N)}$ is its covariance. With this notation, the partition function simplifies to
\begin{align}
    Z^n &= \int [d\tilde\mu(\w^a)][d\cH^{a}_{\mu}] [d\hat H^{a}_{\mu}] \,e^{i\sum_{a,\mu}\hat H^{a}_{\mu} H^{a}_{\mu}} e^{-i\sum_{a,\mu}\hat H^{a}_{\mu} \bm{\cW}^a\cdot \bgamma^{\mu}}
\end{align}
and with the assumption that each sample $\bgamma^\mu$ is drawn i.i.d. from the normal distribution $\cN(0, \bSigma_\bGamma)$, its dataset average over $\bgamma^{\mu}$ yields
\begin{align}
    \braket{Z^n} &= \int [d\tilde\mu(\w^a)] \lrsqpar{\int[d\cH^{a}] [d\hat H^{a}] \,e^{i\sum_{a}\hat H^{a} H^{a}} \Braket{e^{-i\sum_{a}\hat H^{a} \bm{\cW}^a\cdot \bgamma}}_{\bgamma}}^{P}.
\end{align}
We define the order parameters $q^a = \bm{\cW}^a\cdot \bgamma$ and consider their first and second moments:
\begin{align}
    \braket{q^a} = 0,\quad C^{ab} = \braket{q^a q^b} = \bm{\cW}^a \bSigma_\bGamma \bm{\cW}^b,
\end{align}
Then the quenched average over $\bgamma \sim \cN(0, \bSigma_\bGamma)$ gives:
\begin{align}
    \braket{Z^n} &= \int [d C^{ab}]  [d\tilde\mu(\w^a)]  \delta\lrpar{C^{ab} - \bm{\cW}^a \bSigma_\bGamma \bm{\cW}^b} \lrsqpar{\int [d\cH^{a}] [d\hat H^{a}] \,e^{i\sum_{a}\hat H^{a} H^{a}} e^{-\frac{1}{2}\sum_{a,b}\hat H^{a} C^{ab}\hat H^{a}}}^{{P}}\nonumber\\
    &= \int [d C^{ab}]  [d\tilde\mu(\w^a)] \delta\lrpar{C^{ab} - \bm{\cW}^a \bSigma_\bGamma \bm{\cW}^b} \lrsqpar{\int [d\cH^{a}] \,e^{-\frac{1}{2}\sum_{a,b} H^{a} \lrpar{C^{ab}}^{-1} H^{b} -\frac{1}{2}\log\det(C^{ab})}}^{{P}},
\end{align}
where in the last line we performed the integral over $[d\hat H^{a}]$. Furthermore, we integrate in new variables $\hat C^{ab}$ to replace the delta function:
\begin{align}
    \braket{Z^n} &= \int [d C^{ab}] [d\hat C^{ab}]  [d\tilde\mu(\w^a)] e^{i\sum_{ab}\hat C^{ab}\lrpar{C^{ab} - \bm{\cW}^a \bSigma_\bGamma \bm{\cW}^b}} \lrsqpar{\int [d\cH^{a}] \,e^{-\frac{1}{2}\sum_{a,b} H^{a} \lrpar{C^{ab}}^{-1} H^{b} -\frac{1}{2}\log\det(C^{ab})}}^{{P}}\nonumber\\
    & \equiv \int [d C^{ab}] [d\hat C^{ab}] e^{-n\frac{\beta N}{2} (G_0 + \alpha G_1)},
\end{align}
where
\begin{align}
    &G_0= -\frac{2}{n\beta N}\log(\int [d\tilde\mu(\w^a)] e^{i\sum_{ab}\hat C^{ab}\lrpar{C^{ab} - \bm{\cW}^a \bSigma_\bGamma \bm{\cW}^b}}),\nonumber\\
    &[d\tilde\mu(\w^a)] = \prod_a d\mu(\w^a)e^{-\beta {\w^a}^\top \J_{\bpsi\bar\bpsi} \bar\w},\nonumber\\
    &\bm{\cW}^a = \begin{pmatrix} \w^a \\ -\bar\w \end{pmatrix}, \quad \bSigma_\bGamma \equiv \frac{1}{N}\begin{pmatrix}
        \bSigma_\bpsi & \bSigma_{\bpsi\bar\bpsi}\\
        \bSigma_{\bpsi\bar\bpsi}^\top & \bSigma_{\bar\bpsi}
    \end{pmatrix}
\end{align}
and
\begin{align}
    &G_1 = -\frac{2}{n\beta}\log(\int [d\cH^{a}] \,e^{-\frac{1}{2}\sum_{a,b} H^{a} \lrpar{C^{ab}}^{-1} H^{b} -\frac{1}{2}\log\det(C^{ab})}),\nonumber\\
    &[d\cH^{a}] = \prod_a dH^a \, e^{- \frac{\beta}{\lambda} \ell(H^a)}
\end{align}
The coefficients were chosen so that $G_0$ and $G_1$ remain $\cO(1)$ in the thermodynamic limit $N\to\infty$ while $\alpha \equiv P/N \sim \cO(1)$. Next, we compute these two terms under the replica symmetric ansatz:
\begin{align}\label{eq:replica_symmetric_ansatz}
    \C &= \frac{1}{\beta}\lrpar{Q\I_n + \beta C \1_n\1_n^\top} \nonumber\\
    \hat \C &= \frac{{\beta N}}{2i}\lrpar{\hat Q\I_n + \beta \hat C \1_n\1_n^\top}.
\end{align}

\noindent\textbf{The $G_0$ term: }
Representing $\C$ and $\hat\C$ as $n\times n$-matrices with entries $C^{ab}$ and $\hat C^{ab}$, the $G_0$ term can be written as:
\begin{align}
    G_0 &= -\frac{2i}{n\beta N}\Tr\hat\C\C -\frac{2}{n\beta N}\log(\int [d\tilde\mu(\w^a)] e^{-i\sum_{ab}\hat C^{ab}\bm{\cW}^a \bSigma_\bGamma \bm{\cW}^b})
\end{align}
Choosing $d\mu(\w^a) = \lrpar{\frac{\beta}{2\pi}}^{N/2}e^{-\frac{\beta}{2} \norm{\w^a}^2}$, the measure simplifies to
\begin{align}
    [d\tilde\mu(\w^a)] &= \prod_a d\mu(\w^a)e^{-\beta {\w^a}^\top \J_{\bpsi\bar\bpsi} \bar\w} = \lrpar{\frac{\beta}{2\pi}}^{nN/2}\prod_a d\w^a e^{- \frac{\beta}{2}\lrpar{\norm{\w^a}^2 + 2{\w^a}^\top \J_{\bpsi\bar\bpsi} \bar\w}}
\end{align}
Taking the constants out:
\begin{align}
    G_0 &= -\frac{2i}{n\beta N}\Tr\hat\C\C -\frac{2}{n\beta N}\log I_0 \nonumber\\
    I_0 &\equiv \lrpar{\frac{\beta}{2\pi}}^{nN/2}\int [d\w^a] e^{-\frac{\beta}{2}\sum_a \lrpar{\norm{\w^a}^2 + 2{\w^a}^\top \J_{\bpsi\bar\bpsi} \bar\w}-i\sum_{ab}\hat C^{ab}\bm{\cW}^a \bSigma_\bGamma \bm{\cW}^b}.
\end{align}

The integral $I_0$ must be evaluated based on the structure of $\bSigma_\bGamma$ and $\bm{\cW}^a$ which are given above. Then the term $\bm{\cW}^a \bSigma_\bGamma \bm{\cW}^b$ can be expanded as:
\begin{alignat}{2}
    \bm{\cW}^a \bSigma_\bGamma \bm{\cW}^b &  = \w^a \bSigma_{\bpsi} \w^b  + \bar\w\bSigma_{\bar\bpsi}\bar\w - (\w^a+\w^b)\bSigma_{\bpsi\bar\bpsi}\bar\w
\end{alignat}
Then, the exponent in $I_0$ can be written in a compact form by defining:
\begin{align}
    \W &= \begin{pmatrix} \w^1 & \w^2 &\dots &\w^a &\dots &\w^n \end{pmatrix} \nonumber\\
    \bar\W &= \begin{pmatrix} \bar\w & \bar\w &\dots &\bar\w &\dots &\bar\w \end{pmatrix} = \bar\w \otimes \1_n \nonumber\\
    \X &= \frac{2i}{\beta} \bSigma_{\bpsi} \otimes \hat\C, \quad \Y = \frac{2i}{\beta} \bSigma_{\bpsi\bar\bpsi} \otimes\hat\C, \quad \Z = \frac{2i}{\beta} \bSigma_{\bar\bpsi} \otimes\hat\C, \quad \J = \J_{\bpsi\bar\bpsi} \otimes \I_n
\end{align}
yielding
\begin{align}
    -i\sum_{a,b} \hat C^{ab} \bm{\cW}^a \bSigma_\bGamma \bm{\cW}^b = -\frac{\beta}{2}\lrpar{\W^\top\X\W - 2 \W^\top\Y\bar\W + \bar\W^\top\Z\bar\W}.
\end{align}
Plugging this in the integral, we get
\begin{align}
    I_0 = \lrpar{\frac{\beta}{2\pi}}^{nN/2}\int d\W e^{-\frac{\beta}{2}\lrpar{\W^\top\left(\I + \X\right)\W + 2 \W^\top \left(\J - \Y\right) \bar\W + \bar\W^\top \Z \bar\W}}.
\end{align}
The measure also becomes $[d\w^a] = d\W$ and the integral over $nN$-dimensional space can be simply evaluated to:
\begin{align}
    \log I_0 =&  -\frac{1}{2}\log\det(\I + \X) - \frac{\beta}{2} \bar\W^\top \lrsqpar{\Z - \left(\J - \Y\right)^\top\left(\I + \X\right)^{-1}\left(\J - \Y\right)} \bar\W.
\end{align}
Plugging in the replica symmetric ansatz $\hat \C = \frac{{\beta}N}{2i}\left(\hat Q\I_n + \beta \hat C \1_n\1_n^\top\right)$ Eq.\eqref{eq:replica_symmetric_ansatz}, matrices $\X$, $\Y$ and $\Z$ simplify to
\begin{align}
    \X &= \hat Q \bSigma_{\bpsi}\otimes \I_n +  \beta\hat C \bSigma_{\bpsi} \otimes \1_n\1_n^\top,\nonumber\\
    \Y &= \hat Q \bSigma_{\bpsi\bar\bpsi}\otimes \I_n +  \beta\hat C \bSigma_{\bpsi\bar\bpsi} \otimes \1_n\1_n^\top,\nonumber\\
    \Z &= \hat Q \bSigma_{\bar\bpsi}\otimes \I_n +  \beta\hat C \bSigma_{\bar\bpsi} \otimes \1_n\1_n^\top.
\end{align}
The $\log\det(\I + \X)$ simplifies to
\begin{align}
    &\log\det\lrsqpar{(\I_N + \hat Q \bSigma_{\bpsi})\otimes\I_n} + \log\det\lrsqpar{\I + \beta\hat C \bSigma_{\bpsi}(\I_N + \hat Q \bSigma_{\bpsi})^{-1}\otimes\1_n\1_n^\top}\nonumber\\
    &= n\log\det\G + \Tr\log(\I + \beta \hat C \bSigma_{\bpsi} \G^{-1}\otimes\1_n\1_n^\top)\nonumber\\
    &= n\beta N \lrpar{\frac{1}{\beta N}\log\det\G + \hat C \frac{1}{N}\Tr \bSigma_{\bpsi} \G^{-1}} + \cO(n^2)
\end{align}
where we defined  $\G \equiv \I_N + \hat Q \bSigma_{\bpsi}$, used the identity $\log\det\A = \Tr\log\A$ (second line), and Taylor expanded the $\log$ in small $n$ (last line).

Next, we compute $\left(\I + \X\right)^{-1}$ in the limit $n\to 0$:
\begin{align}
    \left(\I + \X\right)^{-1} &= \lrpar{\lrpar{\I_N + \hat Q \bSigma_{\bpsi}}^{-1} \otimes \I_n} \lrpar{\I + \beta\hat C \bSigma_{\bpsi} \G^{-1}\otimes\1_n\1_n^\top}^{-1}\nonumber\\
    &= \lrpar{\G^{-1} \otimes \I_n} \lrpar{\I - \beta\hat C \bSigma_{\bpsi} \G^{-1}\otimes\1_n\1_n^\top + \cO(n)}\nonumber\\
    &= \G^{-1} \otimes \I_n - \beta\hat C\G^{-1} \bSigma_{\bpsi}\G^{-1}\otimes\1_n\1_n^\top + \cO(n).
\end{align}
Note that the second and third terms in $\log I_0$ are of the form $\bar\w\otimes\1_n (\dots)\bar\w\otimes\1_n$, meaning that only terms like $\A\otimes\I_n$ in parentheses survive in the leading order. Therefore, the second term in $\log I_0$ becomes:
\begin{align}
    &-\frac{n\beta}{2} \bar\w^\top\lrsqpar{\hat Q \bSigma_{\bar\bpsi} -\left(\J_{\bpsi\bar\bpsi} - \hat Q \bSigma_{\bpsi\bar\bpsi}\right)^\top \G^{-1} \left(\J_{\bpsi\bar\bpsi} - \hat Q \bSigma_{\bpsi\bar\bpsi}\right)} \bar\w + \cO(n^2)
\end{align}
Then, $\log I_0$ becomes
\begin{align}
    -\frac{2}{n\beta N}\log I_0 = & \frac{1}{\beta N}\log\det\G + \hat C   \tr \bSigma_{\bpsi}\G^{-1} + \tr \bar\w\bar\w^\top \lrsqpar{\hat Q \bSigma_{\bar\bpsi} -\left(\J_{\bpsi\bar\bpsi} - \hat Q \bSigma_{\bpsi\bar\bpsi}\right)^\top \G^{-1} \left(\J_{\bpsi\bar\bpsi} - \hat Q \bSigma_{\bpsi\bar\bpsi}\right)},
\end{align}
where we introduced the normalized trace operation
\begin{align}
    \tr \M = \frac{1}{\dim \M}\Tr\M.
\end{align}
Finally, noting that the term $-\frac{2i}{n\beta N}\Tr\hat\C\C$ reduces to $-\frac{1}{\beta}Q\hat Q - \hat C Q - \hat Q C$, and keeping terms up to $\cO(\J_{\bpsi\bar\bpsi})$, $G_0$ becomes:
\begin{align}
    G_0 =& \frac{1}{\beta}\lrpar{\frac{1}{N}\log\det\G-Q\hat Q} - \hat C (Q - \tr \bSigma_{\bpsi} \G^{-1}) - \hat Q \lrpar{C - \tr\bar\w\bar\w^\top\lrsqpar{\bSigma_{\bar\bpsi} - \bSigma_{\bpsi\bar\bpsi}^\top \lrpar{\hat Q\G^{-1}} \bSigma_{\bpsi\bar\bpsi} }}\nonumber\\
    & + 2\tr \J_{\bpsi\bar\bpsi}\bar\w\bar\w^\top\bSigma_{\bpsi\bar\bpsi}^\top \lrpar{\hat Q\G^{-1}},\nonumber\\
    \G =& \I + \hat Q \bSigma_{\bpsi}
\end{align}
At this point, we can make further simplifications by assuming that $\bSigma_{\bpsi}$ is full-rank and using the identity
\begin{align}
    \hat Q \G^{-1} = (\I - \G^{-1})\bSigma_{\bpsi}^{-1} = \bSigma_{\bpsi}^{-1}(\I - \G^{-1}).
\end{align}
Then, we get
\begin{align}
    G_0 =& - \hat C (Q - \tr \bSigma_{\bpsi} \G^{-1}) - \hat Q \lrpar{C - \tr\bar\w\bar\w^\top\lrpar{\bSigma_{\bar\bpsi} - \bSigma_{\bpsi\bar\bpsi}^\top\bSigma_{\bpsi}^{-1} \bSigma_{\bpsi\bar\bpsi}}} \nonumber\\
    & + \tr\bar\w\bar\w^\top \lrpar{\bSigma_{\bpsi}^{-1}\bSigma_{\bpsi\bar\bpsi}}^\top (\I - \G^{-1})\lrpar{\bSigma_{\bpsi}^{-1} \bSigma_{\bpsi\bar\bpsi} + 2\J_{\bpsi\bar\bpsi}}.
\end{align}
Finally, we define the irreducible error
\begin{align}
    E_\infty &= \tr \bar\w\bar\w^\top \bSigma_{\bar\bpsi} - \tr \w_*{\w_*}^\top \bSigma_{\bpsi} \nonumber\\
    {\w_*} &= \bSigma_{\bpsi}^{-1} \bSigma_{\bpsi\bar\bpsi}\bar\w,
\end{align}
where ${\w_*}$ is the target weights projected on the learnable subspace and ${\w_*}^\top \bSigma_{\bpsi}{\w_*}$ is the maximum explainable variance by the model. With these definitions, the equation for $G_0$ finally simplifies to:
\begin{align}
    G_0 &= -\hat C (Q - \tr \bSigma_{\bpsi} \G^{-1}) - \hat Q \lrpar{C - E_\infty} + \tr \w_*{\w_*}^\top (\I - \G^{-1}) + 2\tr \J_{\bpsi\bar\bpsi}\bar\w{\w_*}^\top (\I - \G^{-1}),\nonumber\\
    \G &= \I + \hat Q \bSigma_{\bpsi},\nonumber\\
    {\w_*} &= \bSigma_{\bpsi}^{-1} \bSigma_{\bpsi\bar\bpsi}\bar\w,\nonumber\\
    E_\infty &= \tr \bar\w\bar\w^\top \bSigma_{\bar\bpsi} - \tr \w_*{\w_*}^\top \bSigma_{\bpsi},
\end{align}

\noindent\textbf{The $G_1$ term: }
The $G_1$ term is given by
\begin{align}
    &G_1 = -\frac{2}{n\beta}\lrpar{-\frac{1}{2}\log\det(\C) + \log I_1},\nonumber\\
    &I_1 = \int [d\cH^{a}] e^{-\frac{1}{2} \sum_{a,b} H^a \left(C^{ab}\right)^{-1} H^b}, \quad [d\cH^{a}] = \prod_a dH^a \, e^{- \frac{\beta}{\lambda} \ell(H^a)}
\end{align}
Under the replica symmetric ansatz $\C = \frac{1}{\beta}\lrpar{Q\I_n + \beta C \1_n\1_n^\top}$:
\begin{align}
    \log\det(\C) &= n\left(\log(Q/\beta) + \frac{\beta C}{Q}\right)+ \cO(n),\nonumber\\
    \C^{-1} &= \frac{\beta}{Q}\I_n - \frac{\beta^2 C}{Q^2}\1_n\1_n^\top + \cO(n),
\end{align}
where we used the identity $\log\det\A = \Tr\log\A$ for the first expression and used the Sherman-Morrison formula for the second expression. Hence, $I_1$ becomes
\begin{align}
    I_1 &= \int [d\cH^a] e^{- \frac{\beta}{2Q} \sum_a (H^a)^2 + \frac{\beta^2 C}{2Q^2}\sum_{a,b}H^a H^b}.
\end{align}
Using the Hubbard-Stratonovich transformation, this integral can be written as
\begin{align}
   I_1(C, C_0) &= \int DT\left[\int d\cH e^{- \frac{\beta}{2Q} H^2 + \frac{\beta\sqrt{C}}{Q} H T} \right]^n\nonumber\\
   &=\int DT\left[e^{\frac{\beta C}{2Q}T^2}\int d\cH  e^{-\frac{\beta}{2Q} (H - \sqrt{C}T)^2} \right]^n\nonumber\\
   &=\int DT\left[e^{\frac{\beta C}{2Q}T^2 + \frac{1}{2}\log(Q/\beta) + \log z(T)}\right]^n,
\end{align}
where $DT = \frac{dT}{\sqrt{2\pi}}e^{-T^2/2}$ is the standard Gaussian measure and we defined
\begin{align}
    z(T) = \int d\cH \sqrt{\frac{\beta}{Q}} e^{-\frac{\beta}{2Q} (H - \sqrt{C}T)^2} = \int dH \sqrt{\frac{\beta C}{Q}} e^{-\frac{\beta C}{2Q} (H - T)^2 - \frac{\beta}{\lambda} \ell(H\sqrt{C})},
\end{align}
where we redefined $H \to H\sqrt{C}$. Using $\int DT A^n \approx \int DT (1 + n\log A) = 1 + n \int DT \log A$ in the limit $n\to 0$, we obtain:
\begin{align}
    \log I_1 = n\frac{\beta C}{2Q} + \frac{n}{2}\log(Q/\beta) + n\int DT \log z(T).
\end{align}
Inserting this expression in $G_1$, we simplify
\begin{align}
    G_1 &= -\frac{2}{n\beta}\lrpar{-\frac{1}{2}\log\det(\C) + \log I_1},\nonumber\\
     &= \frac{1}{\beta}\log(Q/\beta) + \frac{C}{Q} - \lrpar{\frac{C}{Q} + \frac{1}{\beta}\log(Q/\beta) + \frac{2}{\beta}\int DT \log z(T)}\nonumber\\
    &= -\frac{2}{\beta}\int DT \log z(T).
\end{align}
To evaluate this integral, we first express $z(T)$ as
\begin{align}
    z(T) &=  \int dH \sqrt{\frac{\beta C}{Q}} e^{-\beta L(H)},\quad L(H) = \frac{C}{2Q} \lrsqpar{(H - T)^2 + \frac{2Q}{\lambda C}\ell(H\sqrt{C})}.
\end{align}
Note that as $\beta\to\infty$, the integral concentrates around the saddle point $H^* = \argmin_H L(H)$ of $L(H)$. Plugging this saddle point back, $G_1$ becomes
\begin{align}\label{eq:SI_G1_equation}
    G_1 = -\frac{1}{\beta}\log(\frac{\beta C}{Q}) + 2 \Braket{L^*(T)}_T,
\end{align}
where
\begin{align}
    L^*(T) \equiv L\left(H^*(T)\right), \quad H^*(T) = \argmin_H \frac{C}{2Q} \lrsqpar{(H - T)^2 + \frac{2Q}{\lambda C}\ell(H\sqrt{C})}.
\end{align}
We will consider the $\varepsilon$-insensitive loss function:
\begin{align}
    \ell(x) = \frac{1}{n!} \max(0, \abs{x} - \varepsilon)^n.
\end{align}
Then, we have
\begin{align}\label{eq:SI_effective_loss}
    L(H) = \frac{C}{2Q} \lrsqpar{(H - T)^2 + \zeta\frac{2}{n!} \max(0, \abs{H} - \tilde\varepsilon)^n}
\end{align}
where we defined the following effective quantities:
\begin{align}\label{eq:SI_varepsilon_zeta_definition}
    \tilde\varepsilon \equiv \frac{\varepsilon}{\sqrt{C}},\quad \zeta\equiv \frac{Q}{\lambda}C^{\frac{n-2}{2}}.
\end{align}
Next, we solve for the saddle point $H^*$ for general $n$.
\begin{itemize}
    \item \textbf{When $\abs{H} \leq \tilde\varepsilon$}, the loss function does not contribute and the solution is simply:
    \begin{align}
        H^*(T) = T,\quad \abs{H} \leq \tilde\varepsilon
    \end{align}

    \item \textbf{When $\abs{H} > \tilde\varepsilon$}, sign of $T$ determines the sign of $H$. We replace $H = \sign(T)\abs{H}$ and differentiate Eq.\eqref{eq:SI_effective_loss} with respect to $\abs{H}$ to get
    \begin{align}
        \abs{H^*} - \abs{T} + \zeta\frac{1}{(n-1)!} (\abs{H^*} - \tilde\varepsilon)^{n-1} = 0,\quad n \geq 1.
    \end{align}
    Since $\abs{H^*}>\varepsilon$, we instead consider $\abs{H^*} = \tilde\varepsilon + t$ for $t>0$. Then the solution becomes:
    \begin{align}
        H^*(T) = \sign(T)(\tilde\varepsilon + t),\quad \abs{T} > \tilde\varepsilon,
    \end{align}
    where $t$ is the solution of
    \begin{align}
        t + \frac{\zeta}{(n-1)!} t^{n-1} = \abs{T} - \tilde\varepsilon, \quad t \geq 0.
    \end{align}
    Here, $n=0$ is a special case. Recalling $(-1)! = \Gamma(0) = \infty$, its solution is simply $t = \abs{T} - \tilde\varepsilon$.
\end{itemize}
Hence, the saddle point for general $n$ is given by
\begin{align}
    H^*(T) = \begin{cases} 
      T, & \abs{T} \leq \tilde\varepsilon  \\ \\
      \sign(T)(\tilde\varepsilon + t), & \abs{T} > \tilde\varepsilon
   \end{cases},\qquad 
   L(H^*) = \begin{cases}
      0, & \abs{T} \leq \tilde\varepsilon  \\ \\
      \frac{C}{2Q} \lrsqpar{\lrpar{\tilde\varepsilon + t - \abs{T}}^2 + 2\zeta\frac{t^n}{n!}}, & \abs{T} > \tilde\varepsilon
   \end{cases}
\end{align}
where $t$ is the solution of
\begin{align}
    t + \frac{\zeta}{(n-1)!} t^{n-1} = \abs{T} - \tilde\varepsilon, \quad t \geq 0.
\end{align}
Below, we report its solution for $n=0,1,2$.
\begin{itemize}
    \item $n=0$:
    \begin{align}
        t = \begin{cases} 
              0, & \tilde\varepsilon < \abs{T} \leq \tilde\varepsilon+\sqrt{2\zeta}  \\ \\
              \abs{T} - \tilde\varepsilon, & \abs{T} > \tilde\varepsilon+\sqrt{2\zeta}
           \end{cases},\qquad 
       L(H^*) = \begin{cases}
          0, & \abs{T} \leq \tilde\varepsilon  \\ \\
          \frac{C}{2Q}\lrpar{\abs{T}-\tilde\varepsilon}^2, & \tilde\varepsilon < \abs{T} \leq \tilde\varepsilon+\sqrt{2\zeta}\\ \\
          \frac{C\zeta}{Q}, & \abs{T} > \tilde\varepsilon+\sqrt{2\zeta}
       \end{cases}
    \end{align}

    \item $n=1$:
    \begin{align}
        t = \begin{cases} 
              0, & \tilde\varepsilon < \abs{T} \leq \tilde\varepsilon+\zeta \\ \\
              \abs{T} - \tilde\varepsilon - \zeta, & \abs{T} > \tilde\varepsilon+\zeta
           \end{cases},\qquad 
       L(H^*) = \begin{cases}
          0, & \abs{T} \leq \tilde\varepsilon  \\ \\
          \frac{C}{2Q}\lrpar{\abs{T}-\tilde\varepsilon}^2, & \tilde\varepsilon < \abs{T} \leq \tilde\varepsilon+\zeta\\ \\
          \frac{C}{2Q} \lrsqpar{\zeta^2 + 2\zeta(\abs{T} - \tilde\varepsilon - \zeta)}, & \abs{T} > \tilde\varepsilon+\zeta
       \end{cases}
    \end{align}

    \item $n=2$:
    \begin{align}
        t = \frac{\abs{T} - \tilde\varepsilon}{1+\zeta}, \qquad 
       L(H^*) = \begin{cases}
          0, & \abs{T} \leq \tilde\varepsilon  \\ \\
          \frac{C}{2Q} \frac{\zeta}{1+\zeta} \lrpar{\abs{T} - \tilde\varepsilon}^2, & \abs{T} > \tilde\varepsilon
       \end{cases}
    \end{align}

    \item $n\geq 3$: These cases require solving higher-order polynomial equations. However, $\zeta\sim \lambda^{-1}$ is typically large for small regularization $\lambda \approx 0$. In this limit ($\zeta\to \infty$), we have 
    \begin{align}
        t \approx \lrpar{\frac{(n-1)! (\abs{T} - \tilde\varepsilon)}{\zeta}}^{\frac{1}{n-1}},
    \end{align}
    hence, for $\abs{T} > \tilde\varepsilon$, we have
    \begin{align}
        L(H^*) \approx \frac{C}{2Q}\lrpar{\abs{T} - \tilde\varepsilon}^2 + \cO\lrpar{\zeta^{-\frac{1}{n-1}}\, \lrpar{\abs{T} - \tilde\varepsilon}^{\frac{n}{n-1}}}.
    \end{align}
    This implies that the leading order behavior for $n\geq 3$ should be similar to the $n=2$ case. 
\end{itemize}
Now, we can evaluate the average $\Braket{L(H^*)}_T$. First, we introduce the notation
\begin{align}
    \Braket{f(T)}\big|_{a}^{b} \equiv \int_a^b f(T) \frac{e^{-T^2/2}}{\sqrt{2\pi}} dT.
\end{align}
Hence, $\Braket{L(H^*)}_T = \Braket{L(H^*)}\big|_{-\infty}^{\infty}$. Next, we define functions $f(x)$ and $g(x)$ which we will commonly refer to
\begin{align}\label{eq:SI_f_g_functions}
f(x) &= \erfc\lrpar{\frac{x}{\sqrt{2}}}, \quad g(x) = 1 + x^2 - x \frac{\sqrt{\frac{2}{\pi}}e^{-\frac{x^2}{2}}}{\erfc\lrpar{\frac{x}{\sqrt{2}}}}
\end{align}
We also need to evaluate the following integrals:
\begin{align}
    \Braket{(T-x)^2}\bigg|_{x}^{\infty} &= \frac{1}{2} \sqrt{\frac{2}{\pi}}\int_x^\infty (T-x)^2 e^{-T^2/2} dT = \frac{1}{2} f(x)g(x),\nonumber\\
    \Braket{(T-x)}\big|_{x}^{\infty} &= \frac{1}{2} \sqrt{\frac{2}{\pi}}\int_x^\infty (T-x)\, e^{-T^2/2} dT = \frac{1}{2} f(x) \frac{1-g(x)}{x}
\end{align}
Finally, we evaluate $\Braket{L(H^*)}_T$. While we study only the $n=2$ case, here we compute it for all $n=0,1,2$ for reference.
\begin{itemize}
    \item $n=0$:
    \begin{align}
        \Braket{L(H^*)}_T &= \frac{C}{Q}\lrsqpar{\Braket{(T-\tilde\varepsilon)^2}\bigg|_{\tilde\varepsilon}^{\tilde\varepsilon+\sqrt{2\zeta}} + \Braket{2\zeta}\bigg|_{\tilde\varepsilon+\sqrt{2\zeta}}^\infty} = \frac{C}{Q}\lrsqpar{\Braket{(T-\tilde\varepsilon)^2}\bigg|_{\tilde\varepsilon}^{\infty} - \Braket{(T-\tilde\varepsilon)^2 - 2\zeta}\bigg|_{\tilde\varepsilon+\sqrt{2\zeta}}^\infty}\nonumber\\
        &= \frac{C}{Q}\lrsqpar{\Braket{(T-\tilde\varepsilon)^2}\bigg|_{\tilde\varepsilon}^{\infty} - \Braket{\lrpar{T-\tilde\varepsilon-\sqrt{2\zeta}}^2}\bigg|_{\tilde\varepsilon+\sqrt{2\zeta}}^\infty - \sqrt{8\zeta}\Braket{T-\tilde\varepsilon-\sqrt{2\zeta}}\bigg|_{\tilde\varepsilon+\sqrt{2\zeta}}^\infty}\nonumber\\
        &= \frac{C}{2Q}\lrsqpar{f(\tilde\varepsilon)g(\tilde\varepsilon) - f\big(\tilde\varepsilon+\sqrt{2\zeta}\big)g\big(\tilde\varepsilon+\sqrt{2\zeta}\big) - {\sqrt{8\zeta}}\,f\big(\tilde\varepsilon+\sqrt{2\zeta}\big)\frac{1-g\big(\tilde\varepsilon+\sqrt{2\zeta}\big)}{\tilde\varepsilon+\sqrt{2\zeta}} }\nonumber\\
        &= \frac{C}{2Q}\lrsqpar{f(\tilde\varepsilon)g(\tilde\varepsilon) - f\big(\tilde\varepsilon+\sqrt{2\zeta}\big)g\big(\tilde\varepsilon+\sqrt{2\zeta}\big)\lrpar{1 + \frac{\sqrt{8\zeta}}{\tilde\varepsilon+\sqrt{2\zeta}} \frac{1-g\big(\tilde\varepsilon+\sqrt{2\zeta}\big)}{g\big(\tilde\varepsilon+\sqrt{2\zeta}\big)}}}
    \end{align}

    \item $n=1$:
    \begin{align}
        \Braket{L(H^*)}_T &= \frac{C}{Q}\lrsqpar{\Braket{(T-\tilde\varepsilon)^2}\bigg|_{\tilde\varepsilon}^{\tilde\varepsilon+\zeta} + \Braket{\zeta^2 + 2\zeta(\abs{T} - \tilde\varepsilon - \zeta)}\bigg|_{\tilde\varepsilon+\zeta}^\infty}\nonumber\\
        &= \frac{C}{Q}\lrsqpar{\Braket{(T-\tilde\varepsilon)^2}\bigg|_{\tilde\varepsilon}^{\infty} - \Braket{\lrpar{T-\tilde\varepsilon-\zeta}^2}\bigg|_{\tilde\varepsilon+ \zeta}^\infty}\nonumber\\
        &= \frac{C}{2Q}\lrsqpar{f(\tilde\varepsilon)g(\tilde\varepsilon) - f\big(\tilde\varepsilon+\zeta\big)g\big(\tilde\varepsilon+\zeta\big) }
    \end{align}

    \item $n=2$:
    \begin{align}
        \Braket{L(H^*)}_T &= \frac{C}{Q}\frac{\zeta}{1+\zeta}\Braket{(T-\tilde\varepsilon)^2}\bigg|_{\tilde\varepsilon}^{\infty}\nonumber\\
        &= \frac{C}{2Q}\frac{\zeta}{1+\zeta} f(\tilde\varepsilon)g(\tilde\varepsilon)
    \end{align}
\end{itemize}
Note that the standard $\varepsilon$-SVR corresponds to $n=1$ case and can be analyzed in our framework. Here, we focus on $n=2$ case for simplicity. Since we mainly consider very small ridge parameters for SVR, our calculation above demonstrates that all cases behave similarly in the limit $\zeta\to\infty$.

Plugging the result for $n=2$ case in the expression Eq.\eqref{eq:SI_G1_equation} for $G_1$, we get:
\begin{align}
    G_1 =& \frac{C}{Q + \lambda}f(\tilde\varepsilon)g(\tilde\varepsilon) -\frac{1}{\beta}\log(\frac{\beta C}{Q})
\end{align}
where we used the definition from Eq.\eqref{eq:SI_varepsilon_zeta_definition} to get $\zeta = Q/\lambda$ for $n=2$ .

\subsection{Combining all terms}
In the thermodynamic limit $P \to \infty$, the final partition function after the replica symmetric ansatz becomes:
\begin{align}
    \braket{\log Z} &= \lim_{n\to 0}\frac{\braket{Z^n} - 1}{n} =  -\frac{\beta N}{2} S^*,
\end{align}
where $S^*$ is obtained by solving the following saddle point equation
\begin{align}
    S^* =&  \extr_{C, \hat C, Q, \hat Q}\lrsqpar{G_0 + \alpha G_1 + \frac{1}{\beta}\lrpar{\frac{1}{N}\log\det\G-Q\hat Q - \alpha \log(\frac{\beta C}{Q})}} \nonumber\\
    G_0 =& -\hat C (Q - \tr \bSigma_{\bpsi} \G^{-1}) - \hat Q \lrpar{C - E_\infty} + \tr \lrpar{{\w_*}+2\J_{\bpsi\bar\bpsi}\bar\w}{\w_*}^\top (\I - \G^{-1}),\nonumber\\
    G_1 =& \frac{C}{Q + \lambda}f(\tilde\varepsilon)g(\tilde\varepsilon),
\end{align}
where we have the following definitions
\begin{align}
    E_\infty =& \tr \bar\w\bar\w^\top \bSigma_{\bar\bpsi} - \tr \w_*{\w_*}^\top \bSigma_{\bpsi},\nonumber\\
    \G =& \I + \hat Q \bSigma_{\bpsi}, \quad \tilde\varepsilon = \frac{\varepsilon}{\sqrt{C}},\quad 
    {\w_*} = \bSigma_{\bpsi}^{-1} \bSigma_{\bpsi\bar\bpsi}\bar\w
\end{align}

\subsection{Saddle Point Equations}
We next obtain the saddle point equations for $S$. Let us first minimize for $\hat C$:
\begin{itemize}
    \item Minimizing with respect to $\hat C$ gives $\boxed{Q = \tr \bSigma_{\bpsi} \G^{-1}}$.

    \item Minimizing with respect to $\hat Q$ gives:
        \begin{empheq}[box=\widefbox]{align}
          C  =& - \hat C \tr \bSigma_{\bpsi}\G^{-1}\bSigma_{\bpsi}\G^{-1} +  E_\infty + \tr \lrpar{{\w_*}+2 \J_{\bpsi\bar\bpsi}\bar\w}{\w_*}^\top \G^{-1}\bSigma_{\bpsi}\G^{-1}
        \end{empheq}

    \item Minimizing with respect to $C$ gives
        \begin{align}
            \boxed{\hat Q =  \frac{\alpha f(\tilde\varepsilon)}{Q+\lambda}}
        \end{align}

    \item Minimizing with respect to $Q$ gives:  
        \begin{align}
            \boxed{\hat C =  -\frac{\alpha f(\tilde\varepsilon)}{(Q+\lambda)^2}C g(\tilde\varepsilon)}
        \end{align}
        
    \item As a reminder, we had the following definitions
    \begin{align}
    f(\tilde\varepsilon) &=\erfc\lrpar{\frac{\tilde\varepsilon}{\sqrt{2}}}, \quad g(\tilde\varepsilon) = 1+\tilde\varepsilon^2 - \tilde\varepsilon \frac{\sqrt{\frac{2}{\pi}}e^{-\frac{\tilde\varepsilon^2}{2}}}{\erfc\lrpar{\frac{\tilde\varepsilon}{\sqrt{2}}}}, \quad \tilde\varepsilon = \frac{\varepsilon}{\sqrt{C}},\nonumber\\
    \G &= \I + \hat Q \bSigma_{\bpsi}, \quad 
    {\w_*} = \bSigma_{\bpsi}^{-1} \bSigma_{\bpsi\bar\bpsi}\bar\w,\nonumber\\
    E_\infty &= \tr \bar\w\bar\w^\top \bSigma_{\bar\bpsi} - \tr \w_*{\w_*}^\top \bSigma_{\bpsi}
    \end{align}
    
   \item Notice that both $\hat C$ and $\hat Q$ are functions of $C$ and $Q$. Hence, plugging these terms in, we get two coupled self-consistent equations as above. In general, it is challenging to solve these equations analytically.

   \item At the saddle point, we have
   \begin{align}
    S^* &=  \extr_{C, \hat C, Q, \hat Q}\lrsqpar{G_0 + \alpha G_1} \nonumber\\
    G_0 &= \hat Q \lrpar{E_\infty - C} + \tr \lrpar{{\w_*}+2\J_{\bpsi\bar\bpsi}\bar\w}{\w_*}^\top (\I - \G^{-1})\nonumber\\
    &= \hat Q\lrpar{ E_\infty + \tr \lrpar{{\w_*}+2\J_{\bpsi\bar\bpsi}\bar\w}{\w_*}^\top \bSigma_{\bpsi}\G^{-1} -C} \nonumber \\
    G_1 &= \frac{1}{\alpha} \hat Q C g(\tilde\varepsilon)
\end{align}
\end{itemize}

\subsection{Simplifications}
Let us define the following quantities which appear in the solution
\begin{align}
    \tilde\lambda =& \lambda + \tr \bSigma_{\bpsi} \G^{-1}, \quad \tilde\varepsilon = \frac{\varepsilon}{\sqrt{C}}, \quad \tilde \alpha = \alpha f(\tilde\varepsilon),\nonumber\\
    \G =& \I + \frac{\tilde \alpha}{\tilde\lambda} \bSigma_{\bpsi}, \quad \gamma = \frac{\tilde \alpha}{\tilde\lambda^2}  \tr \bSigma_{\bpsi} \G^{-1}\bSigma_{\bpsi} \G^{-1}
\end{align}
Then the saddle point equations are compactly written as
    \begin{align}
        Q =& \tilde\lambda - \lambda,\quad
        \hat Q = \frac{\tilde \alpha}{\tilde\lambda}, \quad \hat C = -\frac{\tilde \alpha}{\tilde\lambda^2} C g(\tilde\varepsilon), \quad
        C = \frac{1}{1- g(\tilde\varepsilon) \gamma} \cW(\tilde\varepsilon, {\tilde \lambda})\nonumber\\
        &\cW(\tilde\varepsilon, {\tilde \lambda}) = E_\infty + \tr \lrpar{{\w_*}+2 \J_{\bpsi\bar\bpsi}\bar\w}{\w_*}^\top \G^{-1}\bSigma_{\bpsi}\G^{-1}
    \end{align}
In terms of these, the action becomes:
\begin{align}
   \braket{\log Z} &= -\frac{\beta N}{2} S,\nonumber\\
   S =& \frac{\tilde \alpha}{\tilde\lambda} \lrsqpar{C(g(\tilde\varepsilon)-1)+ E_\infty +  \tr \lrpar{{\w_*}+2\J_{\bpsi\bar\bpsi}\bar\w}{\w_*}^\top \bSigma_{\bpsi}\G^{-1} }\nonumber\\
   =& \frac{\tilde \alpha}{\tilde\lambda} \lrsqpar{Cg(\tilde\varepsilon)(1-\gamma) + \frac{\tilde \alpha}{\tilde\lambda}\tr \lrpar{{\w_*}+2\J_{\bpsi\bar\bpsi}\bar\w}{\w_*}^\top \bSigma_{\bpsi}\G^{-1}\bSigma_{\bpsi}\G^{-1}}\nonumber\\
   =& \frac{\tilde \alpha}{\tilde\lambda} C g(\tilde\varepsilon)(1-\gamma)
 + \tr \lrpar{{\w_*}+2\J_{\bpsi\bar\bpsi}\bar\w}{\w_*}^\top \lrpar{\I - \G^{-1}}^2
\end{align}
Note that we can compute the following average quantities :
\begin{align}
   -\frac{2}{N}\frac{\partial \braket{\log Z}}{\partial \beta} &= \Braket{\frac{1}{N}\norm{\w}^2 + \frac{2\alpha}{\lambda} E_{tr}} = S \nonumber\\
   \frac{2\lambda}{\beta N} \frac{\partial \braket{\log Z}}{\partial \lambda} &= \Braket{\frac{2\alpha}{\lambda} E_{tr}} = -\lambda \frac{\partial S}{\partial \lambda},
\end{align}
where the average training error is defined as $\frac{1}{P} \sum_{\mu=1}^P \ell\lrpar{\w \cdot \bpsi^\mu - \bar\w\cdot\bar\bpsi^\mu}$

\subsection{Computing Variations}

We have two coupled self-consistent equations in the form of
\begin{align}
    {\tilde \lambda} &= \cF({\tilde \lambda}, \tilde\varepsilon, x) \equiv \lambda + \tr \bSigma_{\bpsi} \G^{-1} \nonumber\\
    \tilde\varepsilon &= \cG({\tilde \lambda}, \tilde\varepsilon, x) \equiv \varepsilon\sqrt{1- g(\tilde\varepsilon) \gamma} \cW(\tilde\varepsilon, {\tilde \lambda})^{-1/2}\nonumber\\
    S &= \cH({\tilde \lambda}, \tilde\varepsilon, x) \equiv \frac{\tilde \alpha}{\tilde\lambda} \lrsqpar{\frac{\varepsilon^2}{\tilde\varepsilon^2}(g(\tilde\varepsilon)-1)+ E_\infty +  \tr \lrpar{{\w_*}+2\J_{\bpsi\bar\bpsi}\bar\w}{\w_*}^\top \bSigma_{\bpsi}\G^{-1}}\nonumber\\
    &\cW(\tilde\varepsilon, {\tilde \lambda}) = E_\infty + \tr \lrpar{{\w_*}+2 \J_{\bpsi\bar\bpsi}\bar\w}{\w_*}^\top \G^{-1}\bSigma_{\bpsi}\G^{-1}
\end{align}
where $x$ collectively denotes any external variable. Their variations with respect to $x$ is given by:
\begin{align}
    \delta {\tilde \lambda} &= \frac{1}{1- \partial_{\tilde \lambda} \cF}\lrpar{\partial_{\tilde\varepsilon}\cF\delta{\tilde\varepsilon} + \partial_x \cF \delta x}\nonumber\\
    \delta{\tilde\varepsilon} &= \frac{1}{1- \partial_{\tilde\varepsilon} \cG}\lrpar{\partial_{\tilde \lambda}\cG\delta{\tilde \lambda} + \partial_x \cG \delta x}\nonumber\\
    \delta S &= \partial_{\tilde \lambda}\cH\delta{\tilde \lambda} + \partial_{\tilde\varepsilon}\cH\delta{\tilde\varepsilon} + \partial_x \cH \delta x.
\end{align}
Solving these equations, we get
\begin{align}
    \delta {\tilde \lambda} &= \frac{(1- \partial_{\tilde\varepsilon} \cG)\partial_x \cF + \partial_{\tilde\varepsilon}\cF \partial_x\cG }{(1- \partial_{\tilde \lambda} \cF)(1- \partial_{\tilde\varepsilon} \cG)-\partial_{\tilde\varepsilon}\cF \partial_{\tilde \lambda} \cG}\delta x\nonumber\\
    \delta{\tilde\varepsilon} &= \frac{(1- \partial_{\tilde \lambda} \cF)\partial_x \cG + \partial_{\tilde \lambda}\cG \partial_x\cF }{(1- \partial_{\tilde \lambda} \cF)(1- \partial_{\tilde\varepsilon} \cG)-\partial_{\tilde\varepsilon}\cF \partial_{\tilde \lambda} \cG}\delta x.
\end{align}
We need to compute $\partial_{\tilde \lambda}\cF$, $\partial_{\tilde\varepsilon}\cF$, $\partial_{\tilde \lambda}\cG$, $\partial_{\tilde\varepsilon}\cG$. First, we compute the derivatives of $\cF$ and $\cH$:
\begin{align}
    \partial_{\tilde \lambda}\cF = \gamma,\quad \partial_{\tilde\varepsilon}\cF = -\frac{f'}{f}{\tilde \lambda}\gamma, \quad \partial_{\tilde \lambda}\cH = -\frac{f}{{\tilde \lambda}^2}\frac{g(1-\gamma)}{1-g\gamma}\cW(\tilde\varepsilon, {\tilde \lambda}), \quad \partial_{\tilde\varepsilon}\cH = -\frac{f'}{{\tilde \lambda}}\frac{g \gamma}{1-g\gamma}\cW(\tilde\varepsilon, {\tilde \lambda}).
\end{align}
Note that:
\begin{align}
    \frac{\partial_{\tilde\varepsilon}\cH}{\partial_{\tilde \lambda}\cH} = -\frac{\partial_{\tilde\varepsilon}\cF}{1-\partial_{\tilde \lambda}\cF} = \frac{f'{\tilde \lambda}}{f}\frac{\gamma}{1-\gamma}
\end{align}
Inserting the variation of $\delta \tilde\lambda$, we get the simplification:
\begin{align}
    \delta S &= \partial_{\tilde \lambda}\cH\lrpar{\delta{\tilde \lambda} -\frac{\partial_{\tilde\varepsilon}\cF}{1-\partial_{\tilde \lambda}\cF} \delta{\tilde\varepsilon}} + \partial_x \cH \delta x = \lrpar{\frac{\partial_{\tilde \lambda}\cH}{1-\gamma}\partial_x \cF + \partial_x\cH}\delta x\nonumber\\
    &= \lrpar{-\frac{f g}{{\tilde \lambda}^2}\frac{\cW(\tilde\varepsilon, {\tilde \lambda})}{1-g\gamma}\partial_x \cF + \partial_x\cH}\delta x = \lrpar{-f(\tilde\varepsilon) g(\tilde\varepsilon) \frac{C}{{\tilde \lambda}^2}\partial_x \cF + \partial_x\cH}\delta x.
\end{align}
The observables from the action hence only depend on $\cF$. This provides a huge simplification for computing first order derivatives of $\braket{\log Z}$.

\subsection{Observables}\label{sec:SI_observables}

Here, we compute the expected values of the observables from the partition function using our calculations above. First, we are interested in the following observables from the partition function:
\begin{align}
    \frac{2\alpha}{\lambda} E_{tr} &= \frac{2\lambda}{\beta N}\frac{\partial}{\partial \lambda} \braket{\log Z} = -\lambda\frac{\partial S}{\partial \lambda} = \frac{\tilde \alpha}{{\tilde \lambda}}C g(\tilde\varepsilon)\frac{\lambda}{{\tilde \lambda}},\nonumber\\
    \frac{1}{N}\braket{\bar\w \w^\top} &= -\frac{1}{\beta N}\frac{\partial}{\partial \J_{\bpsi\bar\bpsi}} \braket{\log Z} = \frac{1}{2}\frac{\partial S}{\partial \J_{\bpsi\bar\bpsi}} = \frac{\tilde\alpha}{\tilde\lambda}\bar\w{\w_*}^\top\bSigma_{\bpsi}\G^{-1} = \frac{1}{N}\lrpar{\I - \G^{-1}}\lrpar{\bSigma_{\bpsi}^{-1} \bSigma_{\bpsi\bar\bpsi}}\bar\w\bar\w^\top
\end{align}
From these observables we also obtain
\begin{align}
    \braket{\w} &= \lrpar{\I - \G^{-1}}\lrpar{\bSigma_{\bpsi}^{-1} \bSigma_{\bpsi\bar\bpsi}}\bar\w = N \frac{\tilde\alpha}{\tilde\lambda}\G^{-1}\bSigma_{\bpsi\bar\bpsi}\bar\w,\nonumber\\
    \frac{1}{N}\braket{\norm{\w}^2} = S - \frac{2\alpha}{\lambda} E_{tr} &= \frac{\tilde \alpha}{\tilde\lambda} Cg(\tilde\varepsilon)\lrpar{1-\gamma-\frac{\lambda}{\tilde \lambda}} + \frac{1}{N}\bar\w^\top \lrpar{\bSigma_{\bpsi}^{-1}\bSigma_{\bpsi\bar\bpsi}}^\top \lrpar{\I - \G^{-1}}^2\lrpar{\bSigma_{\bpsi}^{-1} \bSigma_{\bpsi\bar\bpsi}}\bar\w,\nonumber\\
    \frac{1}{N}\braket{\w\w^\top} - \frac{1}{N}\braket{\w}\braket{\w}^\top &= Cg(\tilde\varepsilon)\frac{\tilde \alpha}{\tilde\lambda^2}\G^{-1}\bSigma_{\bpsi} \G^{-1} 
\end{align}
where we used the identity $\delta \equiv 1-\gamma-\frac{\lambda}{\tilde \lambda} = \frac{\Tr \G^{-1}\bSigma_{\bpsi} \G^{-1}}{\tilde\lambda}$ at the last line.

Finally, we compute the generalization error as
\begin{align}
    E_g &= \Braket{\bE(\w\cdot\bpsi - \bar\w\cdot\bar\bpsi)^2} = \Braket{\w^\top\bSigma_{\bpsi}\w} + \Braket{\bar\w^\top\bSigma_{\bar\bpsi}\bar\w} - 2\Braket{\w^\top\bSigma_{\bpsi\bar\bpsi}\bar\w}\nonumber\\
    &= Cg(\tilde\varepsilon)\gamma + \bar\w^\top\lrsqpar{\bSigma_{\bar\bpsi}-2\bSigma_{\bpsi\bar\bpsi}^\top\lrpar{\I - \G^{-1}}\lrpar{\bSigma_{\bpsi}^{-1} \bSigma_{\bpsi\bar\bpsi}} + \lrpar{\bSigma_{\bpsi}^{-1}\bSigma_{\bpsi\bar\bpsi}}^\top \bSigma_{\bpsi} \lrpar{\I - \G^{-1}}^2\lrpar{\bSigma_{\bpsi}^{-1} \bSigma_{\bpsi\bar\bpsi}}}\bar\w\nonumber\\
    &= Cg(\tilde\varepsilon)\gamma + \bar\w^\top\lrsqpar{\bSigma_{\bar\bpsi}-\bSigma_{\bpsi\bar\bpsi}^\top\lrpar{2\I - \bSigma_{\bpsi}^{-1}(\I - \G^{-1})\bSigma_{\bpsi}}\lrpar{\I - \G^{-1}}\lrpar{\bSigma_{\bpsi}^{-1} \bSigma_{\bpsi\bar\bpsi}}}\bar\w\nonumber\\
    &= Cg(\tilde\varepsilon)\gamma + \bar\w^\top\lrsqpar{\bSigma_{\bar\bpsi}-\bSigma_{\bpsi\bar\bpsi}^\top\lrpar{\I - \G^{-2}}\lrpar{\bSigma_{\bpsi}^{-1} \bSigma_{\bpsi\bar\bpsi}}}\bar\w\nonumber\\
    &= Cg(\tilde\varepsilon)\gamma + \cW(\tilde\varepsilon, {\tilde \lambda}) = C.
\end{align}

\subsection{Alternative Formulation in terms of Center-Variation Covariance}

\noindent We consider the data model where $\bpsi = \bar\bpsi + \bdelta$ and hence
\begin{align}
    \bSigma_{\bpsi} &= \bSigma_{\bar\bpsi} + \bSigma_{\bdelta} + \bSigma_{\bar\bpsi\bdelta} + \bSigma_{\bar\bpsi\bdelta}^\top\nonumber\\
    \bSigma_{\bpsi\bdelta} &= \bSigma_{\bdelta} + \bSigma_{\bar\bpsi\bdelta}\nonumber\\
    \bSigma_{\bpsi\bar\bpsi} &= \bSigma_{\bar\bpsi} + \bSigma_{\bar\bpsi\bdelta} = \bSigma_{\bpsi} - \bSigma_{\bpsi\bdelta}.
\end{align}
We also note that:
\begin{align}
\bSigma_{\bpsi\bar\bpsi}^\top\bSigma_{\bpsi}^{-1}\bSigma_{\bpsi\bar\bpsi} &= \lrpar{\I - \bSigma_\bpsi^{-1} \bSigma_{\bpsi\bdelta}}^\top \bSigma_\bpsi \lrpar{\I - \bSigma_\bpsi^{-1} \bSigma_{\bpsi\bdelta}}
\end{align}
Replacing this back in generalization error
\begin{align}
    \cW(\tilde\varepsilon, {\tilde \lambda}) &= \bar\W^\top \G^{-1}\bSigma_{\bpsi}\G^{-1}\bar\W + \bar\w^\top\bSigma_{\bar\bpsi}\bar\w - \bar\W^\top\bSigma_{\bpsi}\bar\W \nonumber\\
    \bar\W &= \lrpar{\I - \bSigma_\bpsi^{-1} \bSigma_{\bpsi\bdelta}}\bar\w
\end{align}

\subsection{Derivation by \citet{loureiro2021learning}}\label{sec:SI_loureiro}

Here, we show that our results can also be obtained by using the generic formula derived in \cite{loureiro2021learning}. For our loss function $\ell(x) = \frac{1}{2} \max(0, \abs{x} - \varepsilon)^2$, we denote the proximal operator as
\begin{align}
    h_y(x) = \argmin_z \left\{ \ell(z-y) + \frac{1}{2V} (x - z)^2 \right\},
\end{align}
Solving the optimization problem as before, we explicitly get:
\[
h_y(x) =
\begin{cases}
x, & |x - y| \leq \varepsilon \\
\displaystyle \frac{x+Vy}{1 + V} + \sign(x-y)\frac{V\varepsilon}{1 + V}, & \abs{x - y} > \varepsilon
\end{cases}, \quad h'_y(x) =
\begin{cases}
1, & |x - y| \leq \varepsilon \\
\displaystyle \frac{1}{1 + V}, & |x - y| > \varepsilon
\end{cases}
\]
for $h_y(x)$ and its first derivative. 

Let $\bOmega \in \bR^{d\times d}$ and $\bPsi \in \bR^{p\times p}$ denote the covariance matrix of $d$-dimensional student features and $p$-dimensional teacher features, respectively, and $\bPhi \in \bR^{p\times d}$ their cross-covariance. $\btheta_0 \in \bR^{p}$ denotes the teacher weights. We define the ratios $\alpha \equiv n/d$ and $\eta \equiv p/d$ where $n$ is the number of training samples. The self-consistent equations reported by \citet{loureiro2021learning} (see Eq.(2.8)) are
\[
\left\{
\begin{aligned}
V &= \tr \bOmega (\lambda \I + \hat V \bOmega)^{-1},\nonumber\\
m &= \frac{\hat m}{{\sqrt\eta}} \tr (\bPhi^\top\btheta_0\btheta_0^\top\bPhi)(\lambda \I + \hat V \bOmega)^{-1},\nonumber\\
q &= \tr \bOmega(\lambda \I + \hat V \bOmega)^{-2} \lrpar{ \hat m^2\bPhi^\top\btheta_0\btheta_0^\top\bPhi + \hat q \bOmega}
\end{aligned}
\right.
\quad
\left\{
\begin{aligned}
\hat{V} &= \frac{\alpha}{V} \mathbb{E}_{s,h}\left[1 - F'(s,h)\right] \\
\hat{m} &= \frac{1}{\sqrt{\rho \eta}} \frac{\alpha}{V}\mathbb{E}_{s,h} \left[ s F(s,h) - \frac{m}{\sqrt{\rho}} F'(s, h) \right] \\
\hat{q} &= \frac{\alpha}{V^2} \mathbb{E}_{s,h} \left[ \left( \frac{m}{\sqrt{\rho}} s + \sqrt{q - \frac{m^2}{\rho}} h - F(s,h) \right)^2 \right]
\end{aligned}
\right.
\]
where we defined the random functions $F(s, h) = h_{y}\lrpar{x}$ and $F'(s, h) = h'_{y}\lrpar{x}$ with
\begin{align}
    x \equiv \frac{m}{\sqrt{\rho}}s + \sqrt{q - \frac{m^2}{\rho}}h, \quad y \equiv \sqrt{\rho}s, \quad s, h \sim \cN(0, 1).
\end{align}

For $\hat V$, the argument for the expectation value is independent of $s, h$ and can be expressed as an average over the random variable $z \equiv x - y \sim \cN(0, \rho + q - 2m)$. The result is
\begin{align}
    \hat V = \frac{\alpha}{1+V} \Braket{\Theta(\abs{z}- \varepsilon)}_z = \frac{\alpha}{1+V}f\lrpar{\tilde\varepsilon}, \quad \tilde\varepsilon = \frac{\varepsilon}{\sqrt{C}},\quad C \equiv \rho + q - 2m
\end{align}
where $f$ was defined in \eqref{eq:SI_f_g_functions}, and $\tilde\varepsilon$ coincides with the definition of effective tube size.

For $\hat m$, let us first observe that $\mathbb{E}_{s,h} \left[ s F(s,h)\right] = \mathbb{E}_{s,h} \left[\frac{d}{ds}F(s,h)\right]$ using integration by parts. Then, we get
\begin{align}
    \hat m = \frac{1}{\sqrt{\eta}} \frac{\alpha}{1+V} \Braket{\Theta(\abs{z}- \varepsilon)}_z = \frac{\hat V}{\sqrt{\eta}}.
\end{align}

For $\hat q$, using the definition of function $g$ in \eqref{eq:SI_f_g_functions}, we get
\begin{align}
    \hat q = \frac{\alpha}{(1+V)^2}\Braket{(\abs{z}- \varepsilon)^2\Theta(\abs{z}- \varepsilon)}_z = \frac{\alpha}{(1+V)^2}f(\tilde\varepsilon)g(\tilde\varepsilon) C.
\end{align}
Before moving on, we define the effective load as $\tilde\alpha = \alpha f(\tilde\varepsilon)$ and effective regularization as $\tilde\lambda \equiv \lambda (1 + V)$. Then, the self-consistent equation for $V$ becomes equivalent to:
\begin{align}
    \tilde\lambda = \lambda + \tr \bOmega \G^{-1}, \quad \G = \I + \frac{\tilde\alpha}{\tilde\lambda}\bOmega.
\end{align}
In terms of these quantities, self-consistent equations become
\begin{align}
    \hat m &= \frac{\lambda}{\sqrt{\eta}} \frac{\tilde\alpha}{\tilde\lambda}, \quad \hat q = C \lambda^2 \frac{\tilde\alpha}{\tilde\lambda^2} g(\tilde\varepsilon), \quad m = \frac{1}{\eta} \frac{\tilde\alpha}{\tilde\lambda} \tr \G^{-1}(\bPhi^\top\btheta_0\btheta_0^\top\bPhi),\nonumber\\
    q &= \frac{1}{\eta}\frac{\tilde\alpha^2}{\tilde\lambda^2}\tr \bOmega \G^{-2} ( \bPhi^\top\btheta_0\btheta_0^\top\bPhi) +C g(\tilde\varepsilon) \frac{\tilde\alpha}{\tilde\lambda^2}\tr \bOmega^2 \G^{-2}.
\end{align}
Next, we define
\begin{align}
    \gamma \equiv \frac{\tilde\alpha}{\tilde\lambda^2}\tr \bOmega^2 \G^{-2}, \quad \w_* \equiv \frac{1}{\sqrt{\eta}}\bOmega^{-1}\bPhi^\top\btheta_0.
\end{align}
Using the identity $\frac{\tilde\alpha}{\tilde\lambda}\bOmega \G^{-1} = \I - \G^{-1}$, we simplify $m$ and $q$ to
\begin{align}
    m &= \tr \bOmega \w_* \w_*^\top - \tr \G^{-1} \bOmega \w_* \w_*^\top,\nonumber\\
    q &= \tr \bOmega \w_* \w_*^\top - 2\tr \G^{-1} \bOmega \w_* \w_*^\top + \tr \G^{-1} \bOmega \G^{-1} \w_* \w_*^\top + C g(\tilde\varepsilon) \gamma
\end{align}
Noting that $\rho = \tr\bPsi \btheta_0\btheta_0^\top$, we can finally evaluate $C = \rho + q - 2m$ and recover our result
\begin{align}
    C &= C g(\tilde\varepsilon) \gamma + \lrpar{\tr \bPsi \btheta_0\btheta_0^\top -  \tr \bOmega \w_* \w_*^\top} + \tr \G^{-1} \bOmega \G^{-1} \w_* \w_*^\top\nonumber\\
    &= \frac{1}{1 - g(\tilde\varepsilon) \gamma} \lrsqpar{E_\infty +  \tr \G^{-1} \bOmega \G^{-1} \w_* \w_*^\top},
\end{align}
where $E_\infty \equiv \tr \bPsi \btheta_0\btheta_0^\top - \tr \bOmega \w_* \w_*^\top$.

\section{Optimal Learning Rates}\label{sec:SI_optimal_learning_rates}

In this section, we compute the optimal learning rates both for ridge regression by solving for $\partial_\lambda E_g = 0$, and for $\varepsilon$-SVR by solving $\partial_\varepsilon E_g = 0$ using the analytical expression we derived. Since $E_g$ is defined through two self-consistent equations, its derivatives also have to be solved self-consistently. Hence, in order to compute optimal hyperparameters $\lambda$ and $\varepsilon$, we need to compute the variations of $\tilde\lambda$ and $\tilde\varepsilon$.

\subsection{Computing Variations}

Again we start with the two coupled self-consistent equations in the form of
\begin{align}
    {\tilde \lambda} &= \cF({\tilde \lambda}, \tilde\varepsilon, x) \equiv \lambda + \tr \bSigma_{\bpsi} \G^{-1} \nonumber\\
    \tilde\varepsilon &= \cG({\tilde \lambda}, \tilde\varepsilon, x) \equiv \varepsilon\sqrt{1- g(\tilde\varepsilon) \gamma} \cW(\tilde\varepsilon, {\tilde \lambda})^{-1/2}\nonumber\\
    &\cW(\tilde\varepsilon, {\tilde \lambda}) = E_\infty + \tr \w_*{\w_*}^\top \G^{-1}\bSigma_{\bpsi}\G^{-1}
\end{align}
where $x$ collectively denotes any explicit variable, and we omitted the terms with $\J_{\bpsi\bar\bpsi}$. We found earlier that the variations of $\tilde\lambda$ and $\tilde\varepsilon$ are given by
\begin{align}\label{eq:SI_second_order_variation_formula}
    \delta {\tilde \lambda} &= \frac{(1- \partial_{\tilde\varepsilon} \cG)\partial_x \cF + \partial_{\tilde\varepsilon}\cF \partial_x\cG }{(1- \partial_{\tilde \lambda} \cF)(1- \partial_{\tilde\varepsilon} \cG)-\partial_{\tilde\varepsilon}\cF \partial_{\tilde \lambda} \cG}\delta x\nonumber\\
    \delta{\tilde\varepsilon} &= \frac{(1- \partial_{\tilde \lambda} \cF)\partial_x \cG + \partial_{\tilde \lambda}\cG \partial_x\cF }{(1- \partial_{\tilde \lambda} \cF)(1- \partial_{\tilde\varepsilon} \cG)-\partial_{\tilde\varepsilon}\cF \partial_{\tilde \lambda} \cG}\delta x.
\end{align}
and that the derivatives $\partial_{\tilde \lambda}\cF$ and $\partial_{\tilde\varepsilon}\cF$ are given by:
\begin{align}
    \partial_{\tilde \lambda}\cF = \gamma,\quad \partial_{\tilde\varepsilon}\cF = -\frac{f'}{f}{\tilde \lambda}\gamma.
\end{align}
Next, we compute the derivative $\partial_{\tilde\varepsilon}\cG$
\begin{align}
    \partial_{\tilde\varepsilon} \log\cG &= -\frac{\partial_{\tilde\varepsilon}(g(\tilde\varepsilon)\gamma)}{2(1-g(\tilde\varepsilon)\gamma)} - \frac{1}{2}\frac{\partial_{\tilde\varepsilon}\cW(\tilde\varepsilon, {\tilde \lambda})}{\cW(\tilde\varepsilon, {\tilde \lambda})} = -\frac{1}{2(1-g(\tilde\varepsilon)\gamma)}\lrsqpar{\partial_{\tilde\varepsilon}(g(\tilde\varepsilon)\gamma) + \frac{\partial_{\tilde\varepsilon}\cW(\tilde\varepsilon, {\tilde \lambda})}{C}}\nonumber\\
    &= -\frac{1}{2(1-g(\tilde\varepsilon)\gamma)}\lrsqpar{\gamma \frac{\partial_{\tilde\varepsilon}(f(\tilde\varepsilon)g(\tilde\varepsilon))}{f(\tilde\varepsilon)} + f(\tilde\varepsilon)g(\tilde\varepsilon)\partial_{\tilde\varepsilon}\frac{\gamma}{f(\tilde\varepsilon)} + \frac{\partial_{\tilde\varepsilon}\cW(\tilde\varepsilon, {\tilde \lambda})}{C}}.
\end{align}
Here, 
\begin{align}
    \frac{\partial_{\tilde\varepsilon}(f(\tilde\varepsilon)g(\tilde\varepsilon))}{f(\tilde\varepsilon)} = -\frac{2(1-g(\tilde\varepsilon))}{\tilde\varepsilon}, \qquad \partial_{\tilde\varepsilon}\frac{\gamma}{f(\tilde\varepsilon)} &= -\frac{2\gamma}{\tilde\varepsilon}\frac{f'\tilde\varepsilon}{f^2} \frac{\frac{\tilde \alpha}{{\tilde \lambda}} \Tr \M^3}{\Tr \M^2},\nonumber\\
    \partial_{\tilde\varepsilon}\cW(\tilde\varepsilon, {\tilde \lambda}) = -2\tr \w_*{\w_*}^\top \G^{-1}\bSigma_{\bpsi}\G^{-1}\lrpar{\partial_{\tilde\varepsilon}\G}\G^{-1},\qquad \partial_{\tilde\varepsilon}\G &= \frac{f'}{f}\frac{\tilde \alpha}{{\tilde \lambda}}\bSigma_{\bpsi},
\end{align}
where we defined $\M \equiv \bSigma_{\bpsi}\G^{-1}$ to shorten the notation. Inserting these back,
\begin{align*}
    \tilde\varepsilon\partial_{\tilde\varepsilon} \log\cG &= -\frac{\tilde\varepsilon}{2(1-g(\tilde\varepsilon)\gamma)}\lrsqpar{-\frac{2\gamma(1-g(\tilde\varepsilon))}{\tilde\varepsilon} -\frac{2\gamma g(\tilde\varepsilon)}{\tilde\varepsilon}\frac{f'\tilde\varepsilon}{f} \frac{\frac{\tilde \alpha}{{\tilde \lambda}} \tr \M^3}{\tr \M^2} -\frac{2}{\tilde\varepsilon}\frac{f'\tilde\varepsilon}{f}\frac{\frac{\tilde \alpha}{{\tilde \lambda}}\tr\M^2\G^{-1}{\w_*}{\w_*}^\top}{C}}\nonumber\\
    &= \frac{1}{(1-g(\tilde\varepsilon)\gamma)}\lrsqpar{{\gamma(1-g(\tilde\varepsilon))} + \frac{f'\tilde\varepsilon}{f}\frac{\tilde \alpha}{{\tilde \lambda}} \lrpar{\gamma g(\tilde\varepsilon)\frac{\tr \M^3}{\tr \M^2} + \frac{\tr\M^2\G^{-1}{\w_*}{\w_*}^\top}{C}}}\nonumber\\
    &= \frac{\gamma(1-g(\tilde\varepsilon))}{(1-g(\tilde\varepsilon)\gamma)} + \frac{f'\tilde\varepsilon}{f}\frac{\tilde \alpha}{{\tilde \lambda}} \lrpar{\frac{\gamma g(\tilde\varepsilon)}{(1-g(\tilde\varepsilon)\gamma)}\frac{\tr \M^3}{\tr \M^2} + \frac{\tr\M^2\G^{-1}{\w_*}{\w_*}^\top}{(1-g(\tilde\varepsilon)\gamma) C}}\nonumber\\
    &= 1 - \frac{1-\gamma}{(1-g(\tilde\varepsilon)\gamma)} + \frac{f'\tilde\varepsilon}{f}\frac{\tilde \alpha}{{\tilde \lambda}} \lrpar{\frac{\gamma g(\tilde\varepsilon)}{(1-g(\tilde\varepsilon)\gamma)}\frac{\tr \M^3}{\tr \M^2} + \frac{\tr\M^2\G^{-1}{\w_*}{\w_*}^\top}{\cW(\tilde\varepsilon, {\tilde \lambda})}}
\end{align*}
Since $\partial_{\tilde\varepsilon} \log\cG = \frac{\partial_{\tilde\varepsilon} \cG}{\cG} = \frac{\partial_{\tilde\varepsilon} \cG}{\tilde\varepsilon}$, we get:
\begin{align}
    \boxed{\partial_{\tilde\varepsilon} \cG = \frac{\gamma(1-g(\tilde\varepsilon))}{(1-g(\tilde\varepsilon)\gamma)} + \frac{f'\tilde\varepsilon}{f}\frac{\tilde \alpha}{{\tilde \lambda}} \lrpar{\frac{\gamma g(\tilde\varepsilon)}{(1-g(\tilde\varepsilon)\gamma)}\frac{\tr \M^3}{\tr \M^2} + \frac{\tr\M^2\G^{-1}{\w_*}{\w_*}^\top}{\cW(\tilde\varepsilon, {\tilde \lambda})}}}
\end{align}
Next, we compute the derivative $\partial_{\tilde\lambda}\cG$
\begin{align}
    \partial_{{\tilde \lambda}} \log\cG &= -\frac{\partial_{{\tilde \lambda}}(g(\tilde\varepsilon)\gamma)}{2(1-g(\tilde\varepsilon)\gamma)} - \frac{1}{2}\frac{\partial_{{\tilde \lambda}}\cW(\tilde\varepsilon, {\tilde \lambda})}{\cW(\tilde\varepsilon, {\tilde \lambda})} = -\frac{1}{2(1-g(\tilde\varepsilon)\gamma)}\lrsqpar{g(\tilde\varepsilon)\partial_{{\tilde \lambda}}\gamma + \frac{\partial_{{\tilde \lambda}}\cW(\tilde\varepsilon, {\tilde \lambda})}{C}}
\end{align}
Here, we get
\begin{align*}
    \partial_{{\tilde \lambda}}\gamma &= -\frac{2\gamma}{{\tilde \lambda}}\lrpar{1-\frac{\tilde \alpha }{{\tilde \lambda}}\frac{\tr\M^3}{\tr\M^2}}, \quad \partial_{\tilde\lambda}\cW(\tilde\varepsilon, {\tilde \lambda}) = -2\tr \w_*{\w_*}^\top \G^{-1}\bSigma_{\bpsi}\G^{-1}\lrpar{\partial_{\tilde\lambda}\G}\G^{-1}, \quad \partial_{{\tilde \lambda}}\G = -\frac{\tilde \alpha}{{\tilde \lambda}^2}\bSigma_{\bpsi}.
\end{align*}
Inserting these
\begin{align*}
    {\tilde \lambda}\partial_{{\tilde \lambda}} \log\cG &= -\frac{{\tilde \lambda}}{2(1-g(\tilde\varepsilon)\gamma)}\lrsqpar{-\frac{2g(\tilde\varepsilon)\gamma}{{\tilde \lambda}}\lrpar{1-\frac{\tilde \alpha}{{\tilde \lambda}}\frac{\tr\M^3}{\tr\M^2}} + \frac{2\tilde \alpha}{{\tilde \lambda}^2} \frac{\tr\M^2\G^{-1}{\w_*}{\w_*}^\top}{C}} \nonumber\\
    &= -\frac{1}{(1-g(\tilde\varepsilon)\gamma)}\lrsqpar{-g(\tilde\varepsilon)\gamma  +\frac{\tilde \alpha}{{\tilde \lambda}}\lrpar{ g(\tilde\varepsilon)\gamma\frac{\tr\M^3}{\tr\M^2} + \frac{\tr\M^2\G^{-1}{\w_*}{\w_*}^\top}{C}}} \nonumber\\
    &= \frac{g(\tilde\varepsilon)\gamma }{(1-g(\tilde\varepsilon)\gamma)}  - \frac{\tilde \alpha}{{\tilde \lambda}}\lrpar{\frac{g(\tilde\varepsilon)\gamma }{(1-g(\tilde\varepsilon)\gamma)}\frac{\tr\M^3}{\tr\M^2} + \frac{\tr\M^2\G^{-1}{\w_*}{\w_*}^\top}{\cW(\tilde\varepsilon, {\tilde \lambda})}}
\end{align*}
Hence, we have
\begin{align}
    \boxed{\partial_{{\tilde \lambda}} \cG =\frac{\tilde\varepsilon}{{\tilde \lambda}}\lrpar{\frac{g(\tilde\varepsilon)\gamma }{(1-g(\tilde\varepsilon)\gamma)}  - \frac{\tilde \alpha}{{\tilde \lambda}}\lrpar{\frac{g(\tilde\varepsilon)\gamma }{(1-g(\tilde\varepsilon)\gamma)}\frac{\tr\M^3}{\tr\M^2} + \frac{\tr\M^2\G^{-1}{\w_*}{\w_*}^\top}{\cW(\tilde\varepsilon, {\tilde \lambda})}}}}.
\end{align}
Let us focus on the following term
\begin{align}
    &\frac{\tilde \alpha}{{\tilde \lambda}}\lrpar{\frac{g(\tilde\varepsilon)\gamma }{(1-g(\tilde\varepsilon)\gamma)}\frac{\tr\M^3}{\tr\M^2} + \frac{\tr\M^2\G^{-1}{\w_*}{\w_*}^\top}{\cW(\tilde\varepsilon, {\tilde \lambda})}} = \frac{g(\tilde\varepsilon)\gamma }{(1-g(\tilde\varepsilon)\gamma)}\frac{\tr\M^2(\I-\G^{-1})}{\tr\M^2} + \frac{\tilde \alpha}{{\tilde \lambda}}\frac{\tr\M^2\G^{-1}{\w_*}{\w_*}^\top}{\cW(\tilde\varepsilon, {\tilde \lambda})} \nonumber\\
    &= \frac{g(\tilde\varepsilon)\gamma}{(1-g(\tilde\varepsilon)\gamma)} - \lrpar{\frac{g(\tilde\varepsilon)\gamma }{(1-g(\tilde\varepsilon)\gamma)}\frac{\tr\M^2\G^{-1}}{\tr\M^2} - 
 \frac{\tilde \alpha}{{\tilde \lambda}}\frac{\tr\M^2\G^{-1} {\w_*}{\w_*}^\top}{\cW(\tilde\varepsilon, {\tilde \lambda})}}\nonumber\\
    &= \frac{g(\tilde\varepsilon)\gamma}{(1-g(\tilde\varepsilon)\gamma)} - \frac{\gamma}{(1-g(\tilde\varepsilon)\gamma)} \cA(\tilde\varepsilon, {\tilde \lambda})
\end{align}
where we defined.
\begin{align}
    \cA(\tilde\varepsilon, {\tilde \lambda}) &= g(\tilde\varepsilon)\frac{\tr\M^2\G^{-1}}{\tr\M^2} - 
 \frac{\tilde \lambda}{C}\frac{\tr\M^2\G^{-1} {\w_*}{\w_*}^\top}{\tr\M^2}\nonumber\\
 &= \frac{\tr\M^2\G^{-1}}{\tr\M^2}\lrpar{g(\tilde\varepsilon) - 
 \frac{{\tilde \lambda}}{C}\frac{\tr\M^2\G^{-1} {\w_*}{\w_*}^\top}{\tr\M^2\G^{-1}}}
\end{align}
This conveniently simplifies $\partial_{{\tilde \lambda}}\cG$ to:
\begin{align}
    \boxed{\partial_{{\tilde \lambda}} \cG =\frac{\tilde\varepsilon}{{\tilde \lambda}} \frac{\gamma}{(1-g(\tilde\varepsilon)\gamma)} \cA(\tilde\varepsilon, {\tilde \lambda})}
\end{align}
Similarly, $\partial_{{\tilde \varepsilon}}\cG$ becomes:
\begin{align}
    \partial_{\tilde\varepsilon} \cG &= \frac{\gamma(1-g(\tilde\varepsilon))}{(1-g(\tilde\varepsilon)\gamma)} + \frac{f'\tilde\varepsilon}{f}\frac{g(\tilde\varepsilon)\gamma}{(1-g(\tilde\varepsilon)\gamma)} - \frac{f'\tilde\varepsilon}{f} \frac{\gamma}{(1-g(\tilde\varepsilon)\gamma)} \cA(\tilde\varepsilon, {\tilde \lambda})\nonumber\\
    &=  \frac{f'\tilde\varepsilon}{f} \frac{\gamma}{(1-g(\tilde\varepsilon)\gamma)}\lrpar{\frac{f}{f'\tilde\varepsilon}(1-g(\tilde\varepsilon)) + g(\tilde\varepsilon) - \cA(\tilde\varepsilon, {\tilde \lambda})}
\end{align}
Using the identity
\begin{align}
    g = 1 + \tilde\varepsilon^2 + \frac{f'\tilde\varepsilon}{f},
\end{align}
this simplifies $\partial_{\tilde\varepsilon} \cG$ to
\begin{align}
    \boxed{\partial_{\tilde\varepsilon} \cG =  \frac{f'\tilde\varepsilon}{f}\frac{\gamma}{(1-g(\tilde\varepsilon)\gamma)}\lrsqpar{h(\tilde\varepsilon) - \cA(\tilde\varepsilon, {\tilde \lambda})}}
\end{align}
where we defined the function 
\begin{align}
    h(\tilde\varepsilon) \equiv \tilde\varepsilon\lrpar{\tilde\varepsilon + \frac{f'}{f} - \frac{f}{f'}} = \frac{(g(\tilde\varepsilon)-1)^2 - g(\tilde\varepsilon)\tilde\varepsilon^2}{g(\tilde\varepsilon) - 1 - \tilde\varepsilon^2}.
\end{align}
We obtain the denominator in Eq.\eqref{eq:SI_second_order_variation_formula} as
\begin{align}
    1- \partial_{\tilde\varepsilon} \cG - \frac{\partial_{\tilde\varepsilon}\cF \partial_{\tilde \lambda} \cG}{1- \partial_{\tilde \lambda} \cF} = 1 - \frac{f'\tilde\varepsilon}{f}\frac{\gamma}{(1-g(\tilde\varepsilon)\gamma)}\lrsqpar{h(\tilde\varepsilon)  - \frac{\cA(\tilde\varepsilon, {\tilde \lambda})}{1-\gamma}},
\end{align}
Finally we get the variation of $\tilde\varepsilon$ as
\begin{align}\label{eq:SI_effective_epsilon_variation}
    \delta\tilde\varepsilon = \frac{\partial_x \cG}{1- \partial_{\tilde\varepsilon} \cG -\frac{\partial_{\tilde\varepsilon}\cF \partial_{\tilde \lambda} \cG}{1- \partial_{\tilde \lambda} \cF}}\delta x + \frac{\frac{\partial_{\tilde \lambda}\cG}{1- \partial_{\tilde \lambda} \cF} \partial_x\cF }{1- \partial_{\tilde\varepsilon} \cG -\frac{\partial_{\tilde\varepsilon}\cF \partial_{\tilde \lambda} \cG}{1- \partial_{\tilde \lambda} \cF}}\delta x
\end{align}

\subsection{Optimal Ridge Paremeter}

In order to compute the optimal ridge parameter, we need to compute the derivative of generalization error $C$ with respect to $\lambda$ when the tube size is zero. Since
\begin{align}
    \partial_\lambda C = -2C \partial_\lambda \log\tilde\varepsilon,
\end{align}
it suffices to compute $\partial_\lambda \tilde\varepsilon$ using Eq.\eqref{eq:SI_effective_epsilon_variation}
\begin{align}
    \partial_\lambda \log \tilde\varepsilon = \frac{1}{1 - \frac{f'\tilde\varepsilon}{f}\frac{\gamma}{(1-g(\tilde\varepsilon)\gamma)}\lrsqpar{h(\tilde\varepsilon)  - \frac{\cA(\tilde\varepsilon, {\tilde \lambda})}{1-\gamma}}}\frac{1}{{\tilde \lambda}} \frac{\gamma}{(1-g(\tilde\varepsilon)\gamma)} \cA(\tilde\varepsilon, {\tilde \lambda}) = \frac{1}{{\tilde \lambda}} \frac{\gamma}{(1-\gamma)} \cA(\tilde\varepsilon, {\tilde \lambda}),
\end{align}
where we used the fact that $\varepsilon = 0$ in the last equality. Hence, the condition for optimal ridge becomes
\begin{align}
     \partial_\lambda C = -\frac{2C}{{\tilde \lambda}} \frac{\gamma}{(1-\gamma)} \cA(\tilde\varepsilon, {\tilde \lambda}) = 0,
\end{align}
which implies another self-consistent equation for effective ridge parameter
\begin{align}
    {\tilde \lambda}_{opt} =  C\frac{\tr\M^2\G^{-1}}{\tr\M^2\G^{-1} {\w_*}{\w_*}^\top}
\end{align}
replacing the original equation $\tilde\lambda = \lambda + \tr \M$.

\subsection{Optimal Tube Size}

For optimal SVR, we set $\lambda = 0$ and consider the derivative of generalization error with respect to $\varepsilon$:
\begin{align}
    \partial_\varepsilon C = \frac{2C}{\varepsilon} - 2C\partial_\varepsilon \log \tilde\varepsilon,
\end{align}
where $\partial_\varepsilon \log \tilde\varepsilon$ is computed using Eq.\eqref{eq:SI_effective_epsilon_variation}
\begin{align}
    \partial_\varepsilon \log \tilde\varepsilon = \frac{1}{\varepsilon} \frac{1}{1 - \frac{f'\tilde\varepsilon}{f}\frac{\gamma}{(1-g(\tilde\varepsilon)\gamma)}\lrsqpar{h(\tilde\varepsilon)  - \frac{\cA(\tilde\varepsilon, {\tilde \lambda})}{1-\gamma}}},
\end{align}
yielding
\begin{align}
    \partial_\varepsilon C = -2\sqrt{C}\frac{f'}{f}\frac{\gamma}{(1-g(\tilde\varepsilon)\gamma)}\frac{\lrsqpar{h(\tilde\varepsilon)  - \frac{\cA(\tilde\varepsilon, {\tilde \lambda})}{1-\gamma}}}{1 - \frac{f'\tilde\varepsilon}{f}\frac{\gamma}{(1-g(\tilde\varepsilon)\gamma)}\lrsqpar{h(\tilde\varepsilon)  - \frac{\cA(\tilde\varepsilon, {\tilde \lambda})}{1-\gamma}}} = 0.
\end{align}
This implies the self-consistency condition for effective tube size
\begin{align}
    h({\tilde \varepsilon}_{opt}) =  \frac{\cA({\tilde \varepsilon}_{opt}, {\tilde \lambda})}{1-\gamma},
\end{align}
complemented with the self-consistent equation $\tilde\lambda = \tr \M$.

\subsection{Optimal Parameters in \figref{fig:figure4}}

The optimal values of $\lambda$ and $\varepsilon$ change as a function of $\alpha$, hence each point in \figref{fig:figure4} has a different optimal ridge parameter or tube size. Here, we show how the optimal parameters depend on the number of samples.

\begin{figure}
    \centering
    \includegraphics[width=0.9\linewidth]{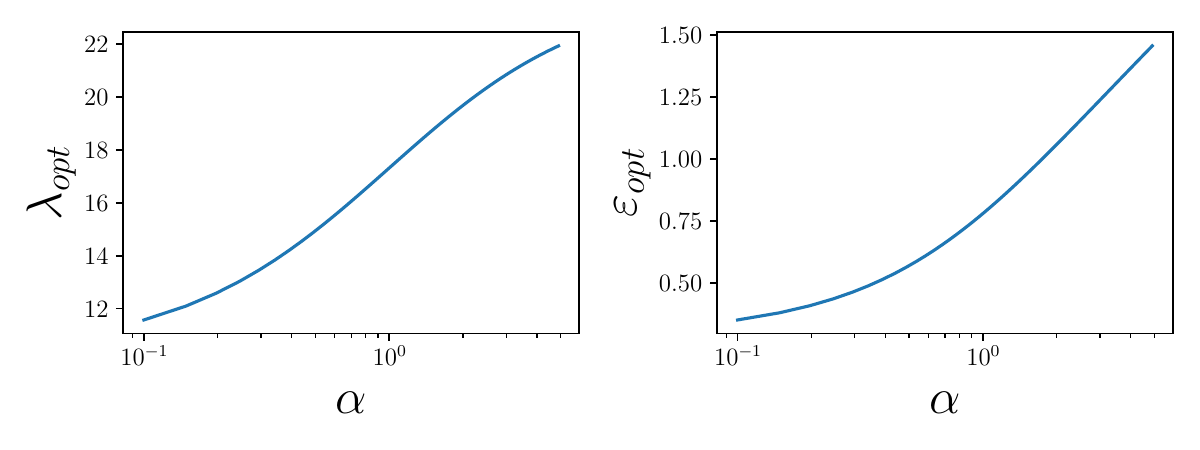}
    \caption{Optimal hyperparameters in \figref{fig:figure4}. Generally, optimal regularization increases with increasing sample sizes.}
    \label{fig:SI_optimal_lambda_epsilon}
\end{figure}

\section{Toy Data Model}\label{sec:SI_toy_data_model}

In order to analyze our result, we consider a toy data model where the center and noise model are given by
\begin{align}
\bpsi &= \bar\bpsi + \sigma \bdelta \nonumber\\
    \bar\bpsi &\sim \cN(0,\I),\quad \bdelta \sim \cN(0,\bSigma_\bdelta)\nonumber\\
    \bSigma_\bdelta &= (1-\beta)\P + \beta (\I - \P) = (1-\beta)\I + \lrpar{2\beta-1} (\I - \P),
\end{align}
where $\P = \I - \frac{1}{N}\bar\w\bar\w^\top$ is the projection matrix to the subspace orthogonal to the coding direction. Here, we assumed that the target weights are normalized such that
\begin{align}
    \Tr \bar\w\bar\w^\top = \Tr \bar\w\bar\w^\top\bSigma_{\bar\bpsi} = \frac{1}{1+\beta\sigma^2}\Tr \bar\w\bar\w^\top\bSigma_{\bpsi} = N 
\end{align}
In this model, the noise features $\bdelta$ are independent of centers $\bar\bpsi$, but their covariance is allowed to correlate with the coding direction $\bar\w$. In this case, we get the following covariance matrices:
\begin{align}
    \bSigma_{\bar\bpsi} &= \bSigma_{\bpsi\bar\bpsi} = \I,\nonumber\\
    \bSigma_{\bpsi} &= \bSigma_{\bar\bpsi} + \sigma^2\bSigma_{\bdelta} = \lrpar{1 + \sigma^2 (1-\beta)}\I + \sigma^2\lrpar{2\beta - 1}\frac{1}{N}\bar\w\bar\w^\top.
\end{align}
Cleaning the last expression, we get $\bSigma_\bpsi$ and its inverse as:
\begin{align}
    \bSigma_\bpsi &= \lrpar{1 + \sigma^2 (1-\beta)}\P + \lrpar{1 + \beta\sigma^2}(\I - \P),\nonumber\\
    \bSigma_\bpsi^{-1} &= \frac{1}{1 + \sigma^2 (1-\beta)}\P + \frac{1}{1 + \beta\sigma^2}(\I - \P)
\end{align}
Hence the error at infinity becomes
\begin{align}
    E_{\infty} &= \tr \bar\w\bar\w^\top \lrpar{\bSigma_{\bar\bpsi} - \bSigma_{\bpsi\bar\bpsi}^\top \bSigma_\bpsi^{-1}\bSigma_{\bpsi\bar\bpsi}} = \frac{\beta\sigma^2}{1+\beta\sigma^2}
\end{align}
We also compute $\G$ and $\G^{-1}$ as:
\begin{align}
    \G &= \lrsqpar{1 + \frac{\tilde\alpha}{\tilde\lambda}\lrpar{1 + \sigma^2 (1-\beta)}}\P + \lrsqpar{1 + \frac{\tilde\alpha}{\tilde\lambda} \lrpar{1 + \beta\sigma^2}}(\I-\P)\nonumber\\
    \G^{-1} &= \frac{1}{1 + \frac{\tilde\alpha}{\tilde\lambda}\lrpar{1 + \sigma^2 (1-\beta)}}\P + \frac{1}{1 + \frac{\tilde\alpha}{\tilde\lambda} \lrpar{1 + \beta\sigma^2}}(\I-\P)
\end{align}
Plugging these into our formula we get:
\begin{align}
    \cF(\tilde \lambda, \tilde\varepsilon) &= \lambda + \frac{1 + \sigma^2 (1-\beta)}{1 + \frac{\tilde\alpha}{\tilde\lambda}\lrpar{1 + \sigma^2 (1-\beta)}} + \frac{1}{N}\lrpar{\frac{1 + \beta\sigma^2}{1 + \frac{\tilde\alpha}{\tilde\lambda}\lrpar{1 + \beta\sigma^2}} - \frac{1 + \sigma^2 (1-\beta)}{1 + \frac{\tilde\alpha}{\tilde\lambda}\lrpar{1 + \sigma^2 (1-\beta)}}},\nonumber\\
    \cG(\tilde \lambda, \tilde\varepsilon) &= \frac{1}{1-g(\tilde\varepsilon)\gamma}\lrpar{\frac{\beta\sigma^2}{1+\beta\sigma^2} + \frac{1}{1+\beta\sigma^2}\frac{1}{\lrpar{1 + \frac{\tilde\alpha}{\tilde\lambda}\lrpar{1 + \beta\sigma^2}}^2}},\nonumber\\
    \gamma &= \frac{\tilde\alpha}{\tilde\lambda^2}\frac{\lrpar{1 + \sigma^2 (1-\beta)}^2}{\lrpar{1 + \frac{\tilde\alpha}{\tilde\lambda}\lrpar{1 + \sigma^2 (1-\beta)}}^2} + \frac{1}{N}\frac{\tilde\alpha}{\tilde\lambda^2}\lrpar{\frac{\lrpar{1 + \beta\sigma^2}^2}{\lrpar{1 + \frac{\tilde\alpha}{\tilde\lambda}\lrpar{1 + \beta\sigma^2}}^2} - \frac{\lrpar{1 + \sigma^2 (1-\beta)}^2}{\lrpar{1 + \frac{\tilde\alpha}{\tilde\lambda}\lrpar{1 + \sigma^2 (1-\beta)}}^2}}.
\end{align}
For $N\gg 1$, we can ignore $\cO(N^{-1})$ terms. Furthermore, we redefine $\tilde\lambda \to \frac{\tilde\lambda}{1 + \sigma^2 (1-\beta)}$ to simplify the self-consistent equations and obtain:
\begin{align}
    \cF(\tilde \lambda, \tilde\varepsilon) &= \frac{\lambda}{1 + \sigma^2 (1-\beta)} + \frac{\tilde\lambda}{\tilde\alpha + \tilde\lambda},\nonumber\\
    \cG(\tilde \lambda, \tilde\varepsilon) &= \frac{1}{1-g(\tilde\varepsilon)\gamma}\lrpar{\frac{\beta\sigma^2}{1+\beta\sigma^2} + \frac{1}{1+\beta\sigma^2}\frac{\tilde\lambda^2}{\lrpar{\tilde\lambda + \tilde\alpha\frac{1 + \sigma^2\beta}{1 + \sigma^2(1-\beta)}}^2}},\nonumber\\
    \gamma &= \frac{\tilde\alpha}{(\tilde\alpha + \tilde\lambda)^2}
\end{align}

\subsection*{Ridgeless limit as exactly solvable model}

The solution to the self-consistent equation for $\tilde\lambda$ in the limit $\lambda \to 0$ becomes
\begin{align}
    \tilde\lambda = 
    \begin{cases}
        1 - \tilde\alpha, & \tilde\alpha < 1\\
        0, & \tilde\alpha \geq 1
    \end{cases},
\end{align}
and the self-consistent equation for $\tilde\varepsilon$ simplifies to:
\begin{align}
    \tilde \varepsilon = \begin{cases}
        \frac{\varepsilon}{\sqrt{E_\infty }} \sqrt{1 - g(\tilde\varepsilon)\tilde\alpha}\lrpar{1 + \frac{1-E_\infty}{E_\infty}\frac{1}{(1 + \frac{\tilde\alpha}{1-\tilde\alpha}\frac{1 + \sigma^2\beta}{1 + \sigma^2(1-\beta)})^2}}^{-1/2} &, \tilde\alpha < 1\\
        \frac{\varepsilon}{\sqrt{E_\infty }} \sqrt{1 - {g(\tilde\varepsilon)}{\tilde\alpha}^{-1}} &, \tilde\alpha \geq 1
    \end{cases}.
\end{align}
Assuming non-zero $E_\infty > 0$, this reduces to the following transcendental equation at the capacity $\tilde\alpha = 1$
\begin{align}\label{eq:SI_trancendental_equation}
    \tilde \varepsilon = d'\sqrt{1 - {g(\tilde\varepsilon)}},
\end{align}
where we defined the discriminability $d' \equiv \frac{\varepsilon}{\sqrt{E_\infty }}$. We asymptotically extract its behavior for large and small $d'$. Note that the range of $\tilde\varepsilon$ is $[0,d')$, since the function $1\geq g(x)>0$. 

For large $d'\gg 1$, $\tilde\varepsilon$ gets large and approaches to $d'$. Expanding the function $g$ in this limit as $g(\tilde\varepsilon) \approx \frac{2}{\tilde\varepsilon^2}$, we get
\begin{align}
    \tilde \varepsilon  \approx d' (1 - d'^{-2}), \quad d' \gg 1.
\end{align}

\begin{figure}[ht]
    \centering
    \includegraphics[width=.4\linewidth]{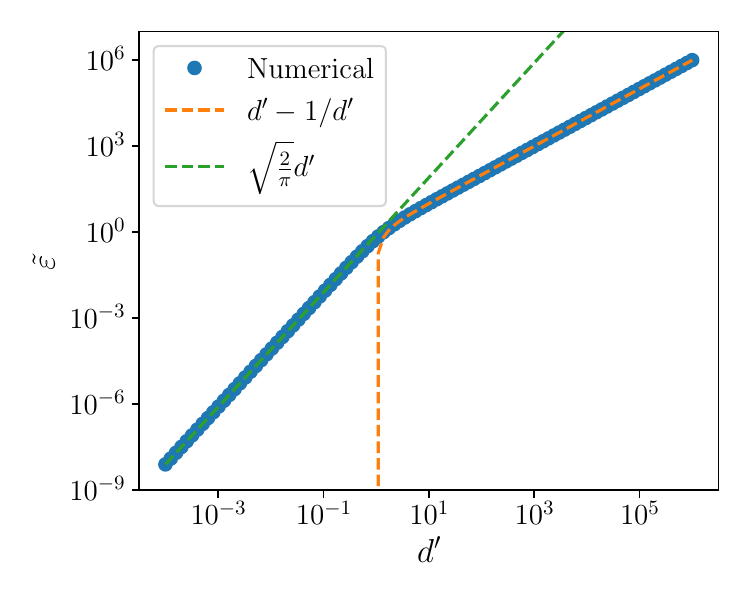}
    \caption{Comparison of the numerical solution to Eq.\eqref{eq:SI_trancendental_equation} to its asymptotic behavior.}
    \label{fig:SI_asymptotic_epsilon}
\end{figure}

Similarly, for small $d'\ll 1$, the argument approaches to zero and admits the expansion $g(\tilde\varepsilon) \approx 1 - \sqrt{\frac{2}{\pi}}\tilde\varepsilon$. Replacing this and squaring both sides, we obtain
\begin{align}
    \tilde \varepsilon  \approx \sqrt{\frac{2}{\pi}} d'^2, \quad d' \ll 1.
\end{align}
In \figref{fig:SI_asymptotic_epsilon}, we confirm the validity of this approximation against its numerical solution.

\section{Spectral Decomposition}\label{sec:SI_spectral_decomposition}

Previously, we had formulated our theory in terms of the covariance matrices of representations where we defined the functions
\begin{align}\label{eq:SI_spectral_self_consistent}
    \cF(\tilde \lambda, \tilde\varepsilon) &= \tr \bSigma_\bpsi \G^{-1}, \;\; \G = \I + \frac{\tilde \alpha}{\tilde\lambda} \bSigma_\bpsi, \;\; \gamma = \partial_{\tilde\lambda}\cF(\tilde \lambda, \tilde\varepsilon) \nonumber\\
    \cG(\tilde \lambda, \tilde\varepsilon) &= \frac{E_\infty + \tr \w_*\w_*^\top \bSigma_\bpsi \G^{-2}}{1- g(\tilde\varepsilon) \gamma}\nonumber\\
    E_{\infty} &= \tr \bar\w\bar\w^\top\bSigma_{\bar\bpsi} -  \tr \w_*\w_*^\top\bSigma_{\bpsi}, \quad \w_* = \bSigma_\bpsi^{-1}\bSigma_{\bpsi\bar\bpsi}\bar\w
\end{align}
where $\tilde \lambda$ and $\tilde \varepsilon$ obey the coupled self-consistent equations given in Eq.\eqref{eq:self_consistent}. Here, we cast our formulas in terms of the spectral properties of representations. Following the kernel notation in \cite{bordelon2020spectrum,canatar2021spectral}, the covariance matrices above are given by:
\begin{align}
    \bSigma_{\bpsi} &= \int \bpsi(\x) \bpsi(\x)^\top d\mu(\x) \nonumber\\
    \bSigma_{\bar\bpsi} &= \int \bar\bpsi(\x) \bar\bpsi(\x)^\top d\mu(\x) \nonumber\\
    \bSigma_{\bpsi\bar\bpsi} &= \int \bpsi(\x) \bar\bpsi(\x)^\top d\mu(\x),
\end{align}
where $d\mu(\x)$ is the distribution of the inputs and we treat representations $\bpsi(\x)$ and $\bar\bpsi(\x)$ as feature maps. We also express the target function and predictor as
\begin{align}
    \bar y(\x) &= \bar\w \cdot \bar\bpsi(\x),\nonumber\\
    y(\x) &= \w \cdot \bpsi(\x).
\end{align}
Note that the asymptotic error is simply the variance of the target function minus the variance of its projection to the learnable subspace:
\begin{align}
    E_\infty = \norm{\bar y(\x)}_2^2 - \norm{ \bar y_\parallel(\x)}_2^2 = \norm{ \bar y_\perp(\x)}_2^2,
\end{align}
which is the unexplainable variance as $\alpha\to\infty$. Note that the explainable portion of the target function can also be obtained by:
\begin{align}
    \bar y_\parallel(\x) = \w_*\cdot\bpsi(\x) = \int d\mu(\x') \, \bar y(\x')\, \lrpar{\bpsi(\x')^\top \bSigma_\bpsi^{-1} \bpsi(\x)},
\end{align}
where the term in the parenthesis can be considered as a projection kernel in the space of square-integrable functions. This allows us to diagonalize our formulas in the basis of $\bSigma_\bpsi$.

We want to study generalization error in regards to task-model alignment and spectral bias \cite{canatar2021spectral} and hence need to obtain the decomposition of the target in the space of square-integrable functions expressable by the feature map $\bpsi(\x)$ via regression. These functions belong to the function space associated with the inner-product kernel $K(\x,\x') = \bpsi(\x)^\top\bpsi(\x')$ and can be diagonalized as
\begin{align}
    K(\x,\x') = \bpsi(\x)^\top\bpsi(\x') = \sum_i \lambda_i \phi_i(\x) \phi_i (\x'); \quad \int \phi_i(\x) \phi_j (\x) d\mu(\x) = \delta_{ij},
\end{align}
where $\{\lambda_i\}$ are non-negative eigenvalues of the kernel and $\{\phi_i(\x)\}$ is a set of complete eigenfunctions with respect to the measure $d\mu(\x)$. While there are infinitely many eigenfunctions with respect to $d\mu(\x)$, there at most $\dim(\bpsi)$ non-zero eigenvalues. We define the index set $\cI^+ = \{i \, | \, \lambda_i > 0\}$ for positive eigenvalues and $\cI^0 = \{i \, | \, \lambda_i = 0\}$ for vanishing eigenvalues. On the other hand, target function generically may have infinitely many components
\begin{align}
    \bar y(\x) = \sum_i \bar a_i \phi_i(\x);\quad \bar a_i = \int \bar y(\x) \phi_i(\x) d\mu(\x),
\end{align}
where $\{\bar a_i\}$ are target coefficients.

Then the equations in Eq.\eqref{eq:SI_spectral_self_consistent} can be written as
\begin{align}
    \cF(\tilde \lambda, \tilde\varepsilon) &= \frac{\tilde\lambda}{N} \sum_{i \in \cI^+} \frac{\lambda_i}{\tilde\lambda + \tilde \alpha \lambda_i}, \;\; \gamma = \frac{1}{N} \sum_{i \in \cI^+} \frac{\tilde \alpha \lambda_i^2}{(\tilde\lambda + \tilde \alpha \lambda_i)^2} \nonumber\\
    \cG(\tilde \lambda, \tilde\varepsilon) &= \frac{1}{1- g(\tilde\varepsilon) \gamma} \lrpar{\sum_{i \in \cI^0} \bar a_i^2 + \sum_{i \in \cI^+} \frac{\tilde\lambda^2}{(\tilde\lambda + \tilde\alpha \lambda_i)^2}\bar a_i^2 },
\end{align}
which exactly matches the equations from \cite{canatar2021spectral} for $\varepsilon = 0$ up to rescaling $\lambda_i \to N \lambda_i$.

This form enables us to analyze the generalization error from an error mode geometry perspective introduced in \cite{canatar2024spectral}. Error mode geometry relates linear predictivity to the geometric properties of representations based on the observation that the generalization error for ridge regression can be decomposed as a sum of mode errors associated with each spectral component \cite{bordelon2020spectrum,canatar2021spectral,canatar2024spectral}. Error modes are defined as
\begin{align}
    W_i(\tilde\alpha) \equiv \frac{\tilde\lambda^2}{1- g(\tilde\varepsilon) \gamma} \frac{\bar a_i^2}{(\tilde\lambda + \tilde\alpha \lambda_i)^2},\quad E_g = \sum_i W_i(\tilde\alpha)
\end{align}
and corresponds to the unexplained variance in the $i^{\text{th}}$ eigendirection. Note that with increasing $\alpha$, the contribution of each mode $\bar a_i^2$ decreases with a rate proportional to the inverse eigenvalue $\lambda_i^{-1}$ associated with that mode, revealing the fact that variances associated with small eigenvalues require more samples to be reduced. Error mode geometry refers to the summarization of these error modes in terms of radius and dimensionality
defined as:
\begin{align}
    R_{em} = \sqrt{\sum_i W_i(\tilde\alpha)^2},\quad D_{em} = \frac{\lrpar{\sum_i W_i(\tilde\alpha)}^2}{\sum_i W_i(\tilde\alpha)^2}
\end{align}
which explicitly relates to the prediction error $E_g = R_{em} \sqrt{D_{em}}$.

\section{Experimental Details}\label{sec:SI_experimental_details}

Experimental details and the code reproducing the figures can be accessed from \url{https://github.com/canatara/svr_code}.

\section{Additional Experiments}\label{sec:SI_additional_experiments}

\subsection{Effect of Target-Noise Correlation}\label{sec:SI_target_noise_correlation}

Even in the presence of noise in the representations, large correlations $\beta$ between noise and the coding direction $\bar\w$ cause more uncertainty in the predictions and, therefore, affect the precision. In \figref{fig:figure4}, we show this effect for various $\beta$, where the noise geometry varies between completely orthogonal and parallel to the coding direction.

\begin{figure}[ht]
    \centering
    \includegraphics[width=.6\linewidth]{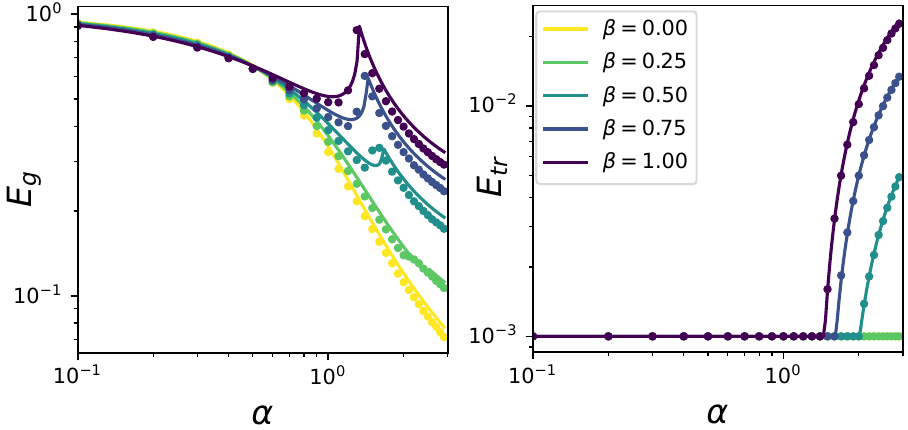}
    \caption{Theoretical (lines) and experimental (dots) learning curves for the toy model with $\sigma=0.5$ and $\epsilon=0.5$. With the same noise strength, increasing the alignment between the noise and coding direction affects the generalization adversely.}
    \label{fig:SI_figure1}
\end{figure}

\subsection{Error Mode Geometry of Real Data}\label{sec:SI_error_mode_additional}
\begin{figure}[ht]
    \centering
    \includegraphics[width=.9\linewidth]{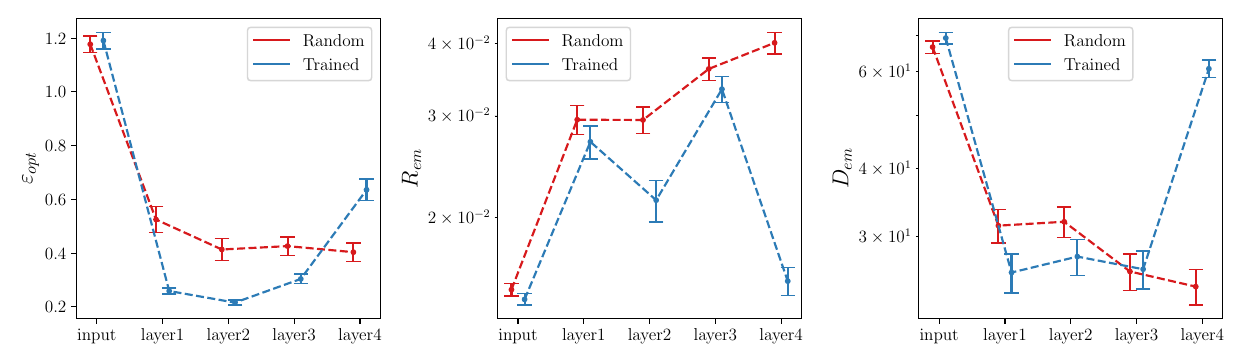}
    \caption{Comparing optimal tube size to error mode geometry \cite{canatar2024spectral}.}
    \label{fig:SI_error_mode}
\end{figure}

\begin{figure}[ht]
    \centering
    \includegraphics[width=.9\linewidth]{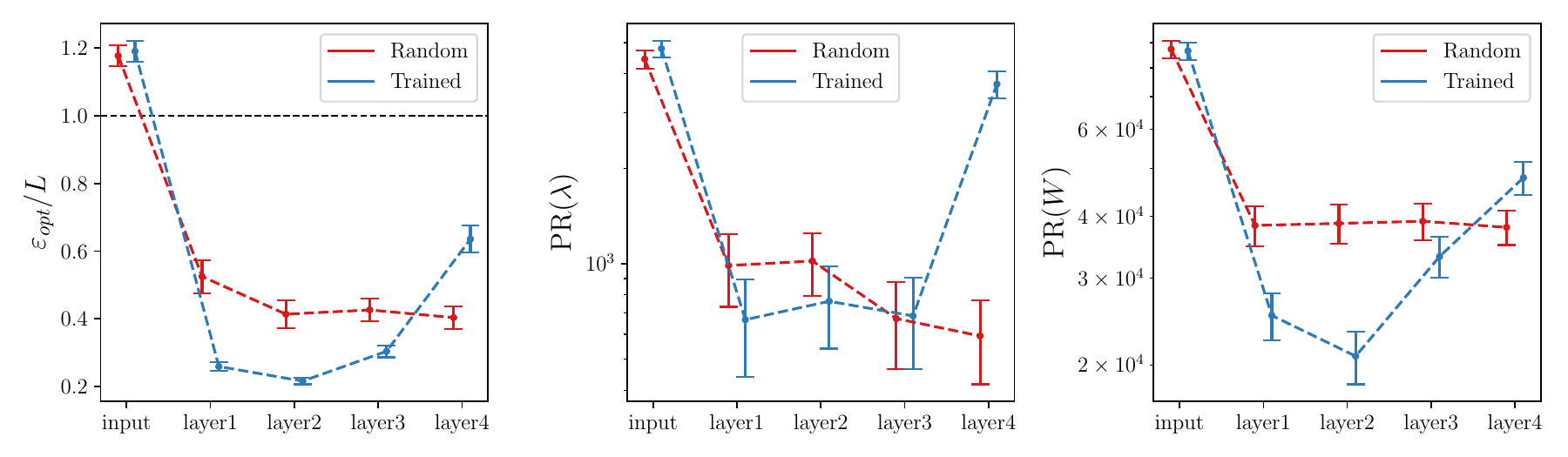}
    \caption{Comparing the optimal tube size to participation ratios of eigenvalues and alignment coefficients.}
    \label{fig:SI_ed_tad}
\end{figure}

Here, the metrics introduced in \cite{canatar2024spectral} and summarized in SI.\ref{sec:SI_spectral_decomposition} are applied to the experiment presented in \figref{fig:figure5}. We show how the error mode geometry decomposition compares to the optimal tube size in Fig.~\figref{fig:SI_error_mode}. From input representation to the first layer, dimensionality $D_{em}$ drives explain lower values of $\varepsilon_{opt}$ despite the increase in $R_{em}$. For random networks, dimensionality across the layer hierarchy decreases, while for the trained networks dimensionality remains the same except for the peak in the last layer. This can be related to the fact that later layers of trained networks are more task-tailored and, hence, become invariant to nuisance variations.
Furthermore, we compare the optimal tube size to the representational properties independent of $\alpha$. We define the participation ratios for the eigenvalues and alignment coefficients as
\begin{align}
    \text{PR}(\lambda) = \frac{\lrpar{\sum_i\lambda_i}^2}{\sum_i \lambda_i^2},\quad \text{PR}(W) = \frac{\lrpar{\sum_i W_i(0)}^2}{\sum_i W_i(0)^2} = D_{em}(0)
\end{align}
which respectively characterize the dimensionality of the neural code and distribution of target components across the model's eigenbasis. In \figref{fig:SI_ed_tad}, we compare optimal tube size to these measures.

\end{document}